\documentstyle[preprint,aps,graphicx,eqsecnum]{revtex}
\newcommand{\be}{\begin{equation}}
\newcommand{\ee}{\end{equation}}
\newcommand{\bea}{\begin{eqnarray}}
\newcommand{\eea}{\end{eqnarray}}

\begin{document}
\draft
\title{{\hbox to\hsize{\hfill KEK-TH-937}}
\vspace{1\baselineskip}
General Formulation for Proton Decay Rate in  
Minimal Supersymmetric SO(10) GUT}
\author{Takeshi Fukuyama$^\dagger$, 
Amon Ilakovac$^\ddagger$, 
Tatsuru Kikuchi$^\dagger$, 
Stjepan Meljanac$^\star$ \\
and Nobuchika Okada$^\diamond$}
\address{$^{\dagger}$Department of Physics, Ritsumeikan University, \\
Kusatsu, Shiga, 525-8577 Japan \\
$^\ddagger$University of Zagreb, Department of Physics, 
P.O. Box 331, Bijeni\v cka cesta 32, HR-10002 Zagreb, Croatia \\
$^\star$Institut Rudjer Bo\v skovi\'c, Bijeni\v cka cesta 54, 
P.O. Box 180, HR-10002 Zagreb, Croatia \\
$^\diamond$Theory Group, KEK, Oho 1-1, Tsukuba, Ibaraki 305-0801, Japan}
\maketitle
\begin{abstract}
We make an explicit formulation for the proton decay rate 
in the minimal renormalizable supersymmetric (SUSY) SO(10) model.  
In this model, the Higgs fields consist of ${\bf 10}$ and ${\bf \overline{126}}$ 
SO(10) representations in the Yukawa interactions with matter 
and of ${\bf 10}$, ${\bf \overline{126}}$, ${\bf 126}$, and ${\bf 210}$ 
representations in the Higgs potential.  
We present all the mass matrices for the Higgs fields 
contained in this minimal SUSY SO(10) model.  
Finally, we discuss the threshold effects of these Higgs fields 
on the gauge couplings unification.   
\end{abstract}
\vspace{0.5cm}
\pacs{PACS number(s): 12.10.-g, 12.10.Dm, 12.10.Kt}
\newpage
\section{Introduction}
Proton decay would be a smoking gun signature for 
Grand Unified Theories (GUTs).  
Unfortunately, no such signal has been seen.  
In fact, very strong experimental limits have been 
set for this process, placing the minimal GUTs 
in a very precarious position.  
SuperKamiokande (SuperK) has set a lower limit on the proton 
lifetime in the channel $p \rightarrow K^{+} \overline{\nu}$ 
as 
\be
\tau (p \rightarrow K^{+} \bar{\nu}) \geq 
2.2 \times 10^{33} \,\, [{\mathrm years}],
\label{SK}  
\ee
at the 90\% confidence level \cite{shiozawa}.  
This has already placed stringent 
constraints on SU(5).  In fact, minimal renormalizable 
SUSY SU(5) model is almost absolutely excluded \cite{murayama}.
\footnote{If we take some textures of the mass matrices 
for fermions and sfermions, we may get a safe region for 
the proton life time in a minimal renormalizable 
SUSY SU(5) model \cite{perez}.  }  
Thus the realistic unified model builders must seriously 
consider the proton life time constraints.  

Now, SO(10) GUTs have been mainly discussed 
in connection with the neutrino oscillations since 
this part reveals the physics beyond the Standard Model.  
In this connection, SO(10) GUTs have some advantages over SU(5) GUTs.  
One of them is that they incorporate the right-handed neutrinos 
as the member of the {\bf 16} dimensional spinor representation 
together with the other standard model fermions, 
and provide the natural explanation of the smallness of 
the neutrino masses through the seesaw mechanism \cite{seesaw}.  
In this paper, we consider the minimal renormalizable SUSY SO(10) model.  
This model contains two Higgs fields ${\bf 10}$ and $\overline{{\bf 126}}$ 
in the Yukawa interactions with matter \cite{babu} \cite{fukuyama}.  
This is a minimal model in the sense that it contains only 
the renormalizable operators at the GUT scale 
and it has the minimal contents of the Higgs fields 
compatible with the low-energy experimental data.  
If we relax the renormalizability at the GUT scale, 
the different minimal SO(10) models are also possible to consider 
\cite{raby} \cite{albright}.  
In this paper, we restrict our arguments within the renormalizable theory 
at the GUT scale, and use the name of minimal in this restricted sense.  
As was shown in \cite{babu} \cite{fukuyama}, this theory is highly predictive.  
However, the recent data \cite{KamLAND} \cite{sno-salt} showed that 
one of our prediction about the neutrino mass square ratio 
$\frac{\Delta m_{\mathrm sol}^2}{\Delta m_{\mathrm atm}^2} \sim 0.19$ 
\cite{fukuyama} is out of $3 \sigma$ allowed region.  
So we need to make re data-fitting compatible with the up to date 
experimental data well.  
However, the development of GUTs and rich experimental data drive us 
to a new stage of precision calculation.  
That is, we must do the precise estimations to include the ambiguities, 
coming from the input data and the threshold corrections.  
The former comes from the fact that there are many input data and 
they have rather large error bars (see, the strange quark mass, for instance).  
We must scan these many possibilities systematically.  
This is tedious but rather technical and is on going in a separate form.  
The latter is more fundamental.  
It depends on the details of the superpotential of the Higgs sector, 
whose effects are not confined in the low-energy data predictions.  
In order to investigate the proton decay rate and the gauge couplings 
unification in precise, we have to determine all the mass spectra of 
the Higgs fields in terms of the parameters presented in this model.  
This is a very complicated task itself and is the main motivation of this work.  
Even in the minimal model, there are so many free parameters.  
So, in practical analysis of the proton decay rate and also the gauge couplings 
unification, one has to reduce the number of free parameters.  
That means we should consider the smallest number of the Higgs contents.  
Thus, we introduce our Higgs system as the most simplest one, 
$\left\{ {\bf 10} \oplus \overline{{\bf 126}} \oplus {\bf 126} 
\oplus {\bf 210} \right\}$.  
The meaning of the introduction of these representations will be 
revealed in the next section.  
Since our results are the general one for the "minimal" renormalizable 
SO(10) models, it can be applicable to any parameter regions.  
For instance, even if we fix the type of the Yukawa couplings in the matter sector 
and also the Higgs potential, the result is not unique.  
If we restrict the values of the parameters in the superpotential 
to some restricted region, we may get the two different types of the seesaw 
mechanism, type-I \cite{babu} \cite {fukuyama} or type-II \cite{rabindra1}.  
In this paper, we do not explain the way to save the model from 
the proton decay rate of SuperK, explicitly.  
Our main purpose of this paper is to produce all the mass spectra of 
the Higgs fields including all the Clebsch-Gordan (CG) coefficients and 
to propose a general formulation which is applicable to any parameter choices.  
In these applications, our theory might be found to be insufficient.  
Even if it is the case, our theory is very useful for more elaborate theory.  

This paper is organized as follows.  
In Sec.II, we give the explicit form of the superpotential in our model.  
In Sec.III, a very brief description of the symmetry breaking procedure 
and the decomposition of the original Higgs fields into the minimal 
supersymmetric standard model (MSSM) are given.  
In Sec.IV, using these techniques, we can get the mass matrices 
for a variety of fields, especially for the would-be Nambu-Goldstone (NG) modes.  
Then we can check that the appropriate NG modes do appear in the mass spectra.  
In Sec.V, we check the mass matrices for the electroweak 
Higgs doublets and consider the conditions for two Higgs doublets 
to remain light.  
In Sec.VI, we derive the formulae for the evaluation 
of the proton decay rate.  
In Sec.VII, we finally check the remaining mass matrices and 
the effects of the threshold corrections on the gauge couplings unification.  
In Appendices, we list up all the coefficients of dimension-five and -six 
operators, which are relevant to proton decay.  
The applications to a more elaborate model will be given in a separate
publication.  
\section{Minimal SO(10) GUT}
In this section, we explain the minimal renormalizable SUSY SO(10) model.  
As mentioned in the introduction, it contains two Higgs 
fields in the Yukawa interactions with matter \cite{babu} \cite{fukuyama}.  
In the SO(10) models, the left- and right-handed 
fermions in a given i-th generation are assigned to 
a single irreducible representation ${\bf 16}_i \equiv \Psi_i$.  
Since ${\bf 16} \otimes {\bf 16} 
= {\bf 10} \oplus {\bf 120} \oplus {\bf 126}$, 
the fermion masses are generated when the Higgs fields 
of ${\bf 10},~{\bf 120}$, and ${\overline{\bf 126}}$ dimensional 
representations develop nonvanishing vacuum expectation values (VEVs).
The use of only one Higgs field, ${\bf 10}$ in the Yukawa interactions 
with matter is obviously ruled out for the description of the realistic 
quark and lepton mass matrices.  
Furthermore, the use of ${\bf \overline{126}}$ dimensional Higgs field 
has desirable properties for providing masses of the right-handed 
Majorana neutrinos.  Also it was found that 
${\bf 10} \, (\equiv H )$ and 
${\bf \overline{126}} \, (\equiv \overline{\Delta} )$ 
are suitable for the mass matrices since they satisfy the 
Georgi-Jarlskog relation.  In order to preserve supersymmetry, 
we must also include the Higgs field 
$\Delta$ of ${\bf 126}$ dimensional representation.  
The Higgs field $\Phi$ of ${\bf 210}$ dimensional representation 
is introduced to break the SO(10) gauge symmetry \cite{clark} 
and to make mix the Higgs doublets included in 
$H$ and ${\overline{\Delta}}$ \cite{babu}.  
Then the minimal Yukawa coupling becomes
\be
W_Y = Y_{10}^{ij} \Psi_i H \Psi_j
+Y_{126}^{ij} \Psi_i \overline{\Delta}\Psi_j,
\label{yukawa}
\ee
and the minimal Higgs superpotential is \cite{clark} 
\cite{lee} \cite{aulakh}  
\be
W=m_1 \Phi^2 + m_2 \Delta \overline{\Delta} 
+m_3 H^2
+\lambda_1 \Phi^3 + \lambda_2 \Phi \Delta \overline{\Delta}
+\lambda_3 \Phi \Delta H + \lambda_4 \Phi \overline{\Delta} H.
\label{lee}
\ee
The interactions of ${\bf 210}$, ${\bf \overline{126}}$, 
${\bf 126}$ and ${\bf 10}$ lead to some complexities 
in decomposing the GUT representations to the MSSM 
and in getting the low energy mass spectra.  
Particularly, the CG coefficients 
corresponding to the decompositions of 
$SO(10) \to SU(3)_C \times SU(2)_L \times U(1)_Y$ have to be found.  
This problem was first attacked by Xiao-Gang He and one of the 
present authors (SM) \cite{he} and further by Lee \cite{lee}.  
But they did not present the explicit form of mass matrices for 
a variety of Higgs fields and also did not perform a formulation 
of the proton life time analysis.  
In this paper we will complete that program 
in the frame of our minimal SO(10) model.  
\section{Symmetry breaking}
In order to discuss the symmetry breaking pattern,  
here we briefly summarize our conventions for the SO(10) indices.  
SO(10) indices $\alpha=1,2,\cdot \cdot \cdot,9,0$ are 
divided into two parts $\alpha=1,2,3,4$ for $SO(4) \cong 
SU(2) \times SU(2)$ 
and $\alpha=5,6,7,8,9,0$ for $SO(6) \cong SU(4)$.  
For the $SO(10) \to SU(3)_C \times SU(2)_L \times U(1)_Y$ decompositions 
it is very useful to define a "$Y$ diagonal basis";  
$1^{\prime}=1 + 2i$, $2^{\prime}=1 - 2i$, 
$3^{\prime}=3 + 4i$, $4^{\prime}=3 - 4i$,  
$5^{\prime}=5 + 6i$, $6^{\prime}=5 - 6i$, 
$7^{\prime}=7 + 8i$, $8^{\prime}=7 - 8i$, 
$9^{\prime}=9 + 0i$, $0^{\prime}=9 - 0i$
(up to normalization factor, $1/\sqrt{2}$).  
Hereafter we use this $Y$ diagonal basis and omit the dashes:  
The ${\bf 10}$ dimensional irreducible representation, $H$ 
is spanned by the states $\alpha=1,2,\cdot \cdot \cdot,9,0$.  
The ${\bf 210}$ dimensional irreducible representation, $\Phi$ and 
the ${\bf \overline{126} \oplus 126}$ dimensional reducible representation 
$\overline{\Delta} + \Delta$, are spanned by the anti-symmetric tensors 
of the fourth rank $\left(\alpha \beta \gamma \delta \right)$ 
and the anti-symmetric tensors of the fifth rank 
$\left(\alpha \beta \gamma \delta \epsilon \right)$, respectively.  
Here and below the bracket $\left(\cdots \right)$ represents 
the total anti-symmetrization of the indices within the bracket.  

The Higgs fields of the minimal $SO(10)$ model contain 
five directions which are singlets 
under $SU(3)_C \times SU(2)_L \times U(1)_Y$. 
The corresponding VEVs are defined by
\bea
\langle \Phi \rangle &=& \sum_{i=1}^3 \phi_i \, \widehat{\phi}_i,
\\
\langle \Delta \rangle &=& v_R \, \widehat{v_R}, \,\,
\langle \overline{\Delta} \rangle = 
\overline{v_R} \, \widehat{\overline{v_R}},
\eea
where $\widehat{\phi}_i \, (i=1,2,3)$ are included in {\bf 210}, 
\bea
\widehat{\phi}_1 &=& -\frac{1}{\sqrt{24}} 
\left(1234 \right), 
\\
\widehat{\phi}_2 &=& -\frac{1}{\sqrt{72}} 
\left(5678 + 5690 + 7890 \right), 
\\
\widehat{\phi}_3 &=& -\frac{1}{\sqrt{144}} 
\left(1256 + 1278 + 1290 + 3456 + 3478 + 3490 \right), 
\eea
$\widehat{{\overline{v_R}}}$ is in ${\bf {\overline{126}}}$ 
\be
\widehat{{\overline{v_R}}} = \frac{1}{\sqrt{120}} 
\left(13579 \right), 
\ee
and $\widehat{v_R}$ is in ${\bf 126}$
\be
\widehat{v_R} = \frac{1}{\sqrt{120}} \left(24680 \right).  
\ee
Notice that 
\bea
\widehat{\phi_i} \cdot \widehat{\phi_j} &=& 
\delta_{ij} \,\, \left(i,j = 1,2,3 \right),
\nonumber\\
\widehat{\overline{v_R}} \cdot \widehat{\overline{v_R}} &=& 
\widehat{v_R} \cdot \widehat{v_R} = 0,
\nonumber\\
\widehat{v_R} \cdot \widehat{\overline{v_R}} &=& 1.  
\eea
Due to the D-flatness condition the absolute values of the VEVs, 
${\overline{v_R}}$ and $v_R$ are equal, 
\be
|\overline{v_R}| = |v_R|. 
\ee
Now we write down the VEV conditions which preserve supersymmetry, 
with respect to the directions 
$\widehat{\phi}_1$, $\widehat{\phi}_2$, $\widehat{\phi}_3$, 
and $\widehat{\overline{v_R}}$, 
respectively.  
\bea
\label{VEV1}
2 m_1 \phi_1 + 3 \lambda_1 \frac{ \phi_3^2}{6 \sqrt{6}} 
+ \lambda_2 \frac{v_R \cdot \overline{v_R}}{10 \sqrt{6}} &=& 0,
\\
\label{VEV2}
2 m_1 \phi_2 + 3 \lambda_1 \left(\frac{\phi_2^2 +\phi_3^2}
{9 \sqrt{2}} \right)
+ \lambda_2 \frac{v_R \cdot \overline{v_R}}{10 \sqrt{2}} &=& 0,
\\
\label{VEV3}
2 m_1 \phi_3 + 3 \lambda_1 \left(\frac{\phi_1 \phi_3}{3 \sqrt{6}}
+\frac{\sqrt{2} \phi_2 \phi_3}{9}\right)
+ \lambda_2 \frac{v_R \cdot \overline{v_R}}{10} &=& 0,
\\
\label{VEV4}
\left\{ m_2 + \lambda_2 \left( \frac{\phi_1}{10 \sqrt{6}}
+ \frac{\phi_2}{10 \sqrt{2}}
+ \frac{\phi_3}{10} \right) \right\} \cdot v_R &=& 0.
\label{VEV}
\eea
Here we consider only the solutions with $|v_R| \neq 0$.
Eliminating $v_R \cdot \overline{v_R}$, 
$\phi_1$ and $\phi_2$ from 
Eqs. (\ref{VEV1})--(\ref{VEV4}), 
one obtains a fourth-order equation in $\phi_3$,  
\be
\label{phi3eq}
\left(\phi_3 + \frac{{\cal M}_2}{10} \right)
\left\{8 \, \phi_3^3 - 15 \, {\cal M}_1 \phi_3^2 
+ 14 \, {\cal M}_1^2 \phi_3 - 3 \, {\cal M}_1 ^3 
+ \left(\phi_3 - {\cal M}_1 \right)^2 {\cal M}_2 \right\} 
= 0,  
\ee
where 
\be
{\cal M}_1 \equiv 12 \, \left(\frac{m_1}{\lambda_1} \right), \, 
{\cal M}_2 \equiv 60 \, \left(\frac{m_2}{\lambda_2} \right).  
\ee
Any solution of the cubic equation in $\phi_3$ is 
accompanied by the solutions 
\bea
\phi_1 &=& - \frac{\phi_3}{\sqrt{6}} 
\frac{\left({\cal M}_1^2 - 5 \, \phi_3^2 \right)}{({\cal M}_1 - \phi_3)^2}, 
\nonumber\\
\phi_2 &=& - \frac{1}{\sqrt{2}} 
\frac{\left({\cal M}_1^2 - 2 \, {\cal M}_1 \phi_3 - \phi_3^2 \right)}{({\cal M}_1 - \phi_3)}, 
\nonumber\\
v_R \cdot \overline{v_R} &=&  \frac{5}{3} \, 
\left(\frac{\lambda_1}{\lambda_2} \right) 
\frac{\phi_3 \left({\cal M}_1 - 3 \, \phi_3 \right) \left({\cal M}_1^2 + \phi_3^2 \right)}
{({\cal M}_1 - \phi_3)^2}.  
\eea
The linear term gives the solution of the fourth-order equation (\ref{phi3eq})
which is very simple, 
$\phi_3 = - 6 \, \left(\frac{m_2}{\lambda_2} \right)$.
It leads to $\phi_1 = -\sqrt{6} \, \left(\frac{m_2}{\lambda_2} \right)$, 
$\phi_2 = -3 \sqrt{2} \, \left(\frac{m_2}{\lambda_2} \right)$    and 
$\sqrt{\left( v_R \cdot \overline{v_R} \right)} = \sqrt{60} \, 
\left(\frac{m_2}{\lambda_2} \right) \sqrt{2 \left(\frac{m_1}{m_2} \right) 
- 3 \left(\frac{\lambda_1}{\lambda_2} \right)}$.  
This solution preserves the $SU(5)$ symmetry.  
Therefore, it is physically not interesting.  
The cubic term solutions lead to the true 
$SU(3)_C \times SU(2)_L \times U(1)_Y$ symmetry.  
\section{Would-be NG bosons}
In order to check the number of NG modes 
we write down the mass matrices for the Higgs(ino) fields which transmute 
the non-MSSM $SO(10)$ gauge fields into very massive gauge fields.  
At first, we list the quantum numbers of 
the would-be NG modes under $SU(3)_C \times SU(2)_L \times U(1)_Y$.  

\begin{itemize}
\item
$
\left[
\left({\bf \overline{3}, 2},\frac{5}{6} \right)
\oplus
\left({\bf 3, 2}, -\frac{5}{6} \right)
\right],
$
\item 
$
\left[
\left({\bf \overline{3}, 2},-\frac{1}{6} \right)
\oplus
\left({\bf 3, 2},\frac{1}{6} \right)
\right],
$
\item
$
\left[
\left({\bf \overline{3}, 1},-\frac{2}{3} \right)
\oplus
\left({\bf 3, 1},\frac{2}{3} \right)
\right], 
$
\item
$
\left[
\left({\bf 1, 1},1 \right)
\oplus
\left({\bf 1, 1},-1 \right)
\right],
$
\item
$\left[\left({\bf 1, 1},0 \right)\right].$
\end{itemize}
Total number of the NG degrees of freedom is :  
${\bf 12 + 12 + 6 + 2 + 1 =  33}$. 
In the following subsections we give explicit expressions for 
the mass matrices and check that their determinants are zero.  
The mass matrices receive  contributions from the 
$F$ terms in the Higgs potential.  
The matrix elements of the mass matrices comprise the 
CG coefficients which appear as coefficients of the triple products of the 
$SU(3)_C \times SU(2)_L \times U(1)_Y$ components of the Higgs superfields.  
For the calculation of the CG coefficients, 
one must first find the explicit expressions for the 
$SU(3)_C \times SU(2)_L \times U(1)_Y$ 
components of the Higgs superfields.  
We will publish the complete tables of the CG coefficients 
of a more general Higgs sector in a separate publication 
and we will list only the mass matices in this paper.  

Note that the mass matrix for every irreducible representation 
under $SU(3)_C \times SU(2)_L \times U(1)_Y$ with $Y \neq 0$ 
and the mass matrix for the corresponding complex conjugate 
representation are equal up to transposition. 
Therefore, only one of the two accompanied mass matrices is listed.  
Of course, when enumerating the total degrees of freedom, 
one has to be careful to include all the mass eigenvalues ($472$ in total).
The mass matrices define the mass part of the superpotential as a 
bilinear form of the fields and corresponding complex conjugate fields. 
The basis for the mass matrix is defined as a row of the fields multiplying
the mass matrix form the left.
\subsection{
$
\left[
\left({\bf \overline{3}, 2},\frac{5}{6} \right)
\oplus
\left({\bf 3, 2}, -\frac{5}{6} \right)
\right]
$}

In the basis $\left\{ \Phi_{\bf(6,2,2)}^{({\bf 3,2},-\frac{5}{6})},
\Phi_{\bf(10,2,2)}^{({\bf 3,2},-\frac{5}{6})} \right\}$ 
(here and hereafter the lower indicies indicate 
$SU(4)_C \times SU(2)_L \times SU(2)_R$ and the upper  
$SU(3)_C \times SU(2)_L \times U(1)_Y$ in the case of double indices), 
the mass matrix is written as  
\be
\left(
\begin{array}{cc}
\begin{array}{c}
2 m_1 - \frac{\lambda_1 \phi_3}{6} \\
\frac{\lambda_1 \phi_3}{3 \sqrt{2}} 
\end{array}
&\begin{array}{c}
\frac{\lambda_1 \phi_3}{3 \sqrt{2}} \\
2 m_1 + \frac{\lambda_1 \phi_2}{3 \sqrt{2}}
-\frac{\lambda_1 \phi_3}{6} 
\end{array}
\end{array}
\right).
\ee
This determinant is indeed zero assuming the VEV 
conditions, Eqs. (\ref{VEV1})--(\ref{VEV4}).  
\subsection{
$
\left[
\left({\bf \overline{3}, 2},-\frac{1}{6} \right)
\oplus
\left({\bf 3, 2},\frac{1}{6} \right)
\right]
$
}   

In the basis $\left\{ \Phi_{\bf(6,2,2)}^{({\bf 3,2},\frac{1}{6})},
\Phi_{\bf(10,2,2)}^{({\bf 3,2},\frac{1}{6})},
\Delta_{\bf(15,2,2)}^{({\bf 3,2},\frac{1}{6})},
\overline{\Delta}_{\bf(15,2,2)}^{({\bf 3,2},\frac{1}{6})} \right\}$, 
the mass matrix is written as   
\be
\left(
\begin{array}{cccc}
\begin{array}{c}
2 m_1 + \frac{\lambda_1 \phi_3}{6} \\
\frac{\lambda_1 \phi_3}{3 \sqrt{2}} \\
-\frac{\lambda_2 \overline{v_R}}{10 \sqrt{3}} \\
0 
\end{array}
&\begin{array}{c}
\frac{\lambda_1 \phi_3}{3 \sqrt{2}} \\
2 m_1 + \frac{\lambda_1 \phi_2}{3 \sqrt{2}} 
+\frac{\lambda_1 \phi_3}{6} \\
-\frac{\lambda_2 \overline{v_R}}{5 \sqrt{6}} \\
0 
\end{array}
&\begin{array}{c}
-\frac{\lambda_2 v_R}{10 \sqrt{3}} \\
-\frac{\lambda_2 v_R}{5 \sqrt{6}} \\
m_2 + \frac{\lambda_2 \phi_2}{30 \sqrt{2}}
+ \frac{\lambda_2 \phi_3}{20} \\
0 
\end{array}
&\begin{array}{c}
0 \\
0 \\
0 \\
m_2 + \frac{\lambda_2 \phi_2}{30 \sqrt{2}}
+ \frac{\lambda_2 \phi_3}{60} 
\end{array}
\end{array}
\right).  
\label{sub}
\ee
This determinant is also equal zero assuming the VEV conditions.  
\subsection{
$
\left[
\left({\bf \overline{3}, 1},-\frac{2}{3} \right)
\oplus
\left({\bf 3, 1},\frac{2}{3} \right)
\right]
$
}

In the basis $\left\{ \Phi_{\bf(15,1,1)}^{({\bf 3,1},\frac{2}{3})},
\Phi_{\bf(15,1,3)}^{({\bf 3,1},\frac{2}{3})},
\overline{\Delta}_{\bf(10,1,3)}^{({\bf 3,1},\frac{2}{3})}
\right\}$,  
the mass matrix is written as  
\be
\left(
\begin{array}{ccc}
\begin{array}{c}
2 m_1 + \frac{\lambda_1 \phi_2}{3 \sqrt{2}} \\
\frac{\lambda_1 \phi_3}{3 \sqrt{2}} \\
-\frac{\lambda_2 v_R}{10 \sqrt{3}}  
\end{array}
&\begin{array}{c}
\frac{\lambda_1 \phi_3}{3 \sqrt{2}} \\
2 m_1 + \frac{\lambda_1 \phi_1}{\sqrt{6}} 
+\frac{\lambda_1 \phi_2}{3 \sqrt{2}} \\
-\frac{\lambda_2 v_R}{5 \sqrt{6}} 
\end{array}
&\begin{array}{c}
-\frac{\lambda_2 \overline{v_R}}{10 \sqrt{3}} \\
-\frac{\lambda_2 \overline{v_R}}{5 \sqrt{6}} \\
m_2 + \frac{\lambda_2 \phi_1}{10 \sqrt{6}} 
+\frac{\lambda_2 \phi_2}{30 \sqrt{2}}
+\frac{\lambda_2 \phi_3}{30} 
\end{array}
\end{array}
\right).  
\ee
This determinant is also equal zero assuming the VEV conditions.  
\subsection{$
\left[
\left({\bf 1, 1},1 \right)
\oplus
\left({\bf 1, 1},-1 \right)
\right] $}

In the basis $\left\{ \Phi_{\bf(15,1,3)}^{({\bf 1,1},1)}, 
\Delta_{\bf(\overline{10},1,3)}^{({\bf 1,1},1)} \right\}$, 
the mass matrix is written as  
\be
\left(
\begin{array}{cc}
\begin{array}{c}
2 m_1 + \frac{\lambda_1 \phi_1}{\sqrt{6}} 
+ \frac{\sqrt{2} \lambda_1 \phi_2}{3} \\
-\frac{\lambda_2 \overline{v_R}}{10} 
\end{array}
&\begin{array}{c}
-\frac{\lambda_2 v_R}{10} \\
m_2 + \frac{\lambda_2 \phi_1}{10 \sqrt{6}}
+\frac{\lambda_2 \phi_2}{10 \sqrt{2}} 
\end{array}
\end{array}
\right).
\ee
This determinant is also equal zero assuming the VEV conditions.  
\subsection{$\left[\left({\bf 1, 1},0 \right)\right]$}
In the basis  $\left\{ 
\Phi_{\bf(1,1,1)}^{({\bf 1,1},0)},
\Phi_{\bf(15,1,1)}^{({\bf 1,1},0)}, 
\Phi_{\bf(15,1,3)}^{({\bf 1,1},0)}, 
\overline{\Delta}_{\bf(10,1,3)}^{({\bf 1,1},0)},
\Delta_{\bf(\overline{10},1,3)}^{({\bf 1,1},0)} \right\}$, 
the mass matrix is written as   
\be 
\left(
\begin{array}{ccccc}
\begin{array}{c}
2 m_1 \\
0 \\
\frac{\lambda_1 \phi_3}{\sqrt{6}} \\
-\frac{\lambda_2 v_R}{10 \sqrt{6}} \\
-\frac{\lambda_2 \overline{v_R}}{10 \sqrt{6}} 
\end{array}
&\begin{array}{c}
0 \\
2 m_1 + \frac{\sqrt{2} \lambda_1 \phi_2}{3} \\
\frac{\sqrt{2} \lambda_1 \phi_3}{3} \\
-\frac{\lambda_2 v_R}{10 \sqrt{2}} \\
-\frac{\lambda_2 \overline{v_R}}{10 \sqrt{2}} 
\end{array}
&\begin{array}{c}
\frac{\lambda_1 \phi_3}{\sqrt{6}}  \\
\frac{\sqrt{2} \lambda_1 \phi_3}{3}  \\
2 m_1 + \frac{\lambda_1 \phi_1}{\sqrt{6}}
+\frac{\sqrt{2} \lambda_1 \phi_2}{3}  \\
-\frac{\lambda_2 v_R}{10} \\
-\frac{\lambda_2 \overline{v_R}}{10} 
\end{array}
&\begin{array}{c}
-\frac{\lambda_2 \overline{v_R}}{10 \sqrt{6}} \\
-\frac{\lambda_2 \overline{v_R}}{10 \sqrt{2}} \\
-\frac{\lambda_2 \overline{v_R}}{10} \\
e4 \\
0 
\end{array}
&\begin{array}{c}
-\frac{\lambda_2 v_R}{10 \sqrt{6}} \\
-\frac{\lambda_2 v_R}{10 \sqrt{2}} \\
-\frac{\lambda_2 v_R}{10} \\
0 \\ 
e4
\end{array}
\end{array}
\right).  
\label{sub} 
\ee
Here $e4 \equiv m_2 + \lambda_2 ( \frac{\phi_1}{10 \sqrt{6}}
+ \frac{\phi_2}{10 \sqrt{2}}
+ \frac{\phi_3}{10})$ 
is nothing but Eq. (\ref{VEV}) 
and Eq. (\ref{sub}) has one zero eigenvalue.   
\section{Electroweak Higgs doublet}
In the standard picture of the electroweak symmetry breaking, 
we have the Higgs doublets which give masses to the matter.  
These masses should be less than or equal to the electroweak scale.  
Since we approximate the electroweak scale as zero, 
we must impose a constraint that the mass matrix should 
have one zero eigenvalue.  

We define 
\be
H_u^{10} \equiv H_{\bf(1,2,2)}^{({\bf 1,2},\frac{1}{2})}, \,
\overline{\Delta}_u \equiv 
\overline{\Delta}_{\bf(15,2,2)}^{({\bf 1,2},\frac{1}{2})}, \, 
\Delta_u \equiv 
\Delta_{\bf(15,2,2)}^{({\bf 1,2},\frac{1}{2})}, \,
\Phi_u \equiv 
\Phi_{\bf(\overline{10},2,2)}^{({\bf 1,2},\frac{1}{2})}.
\ee
and 
\be
H_d^{10} \equiv H_{\bf(1,2,2)}^{({\bf 1,2},-\frac{1}{2})}, \,
\overline{\Delta}_d \equiv 
\overline{\Delta}_{\bf(15,2,2)}^{({\bf 1,2},-\frac{1}{2})}, \,
\Delta_d \equiv \Delta_{\bf(15,2,2)}^{({\bf 1,2},-\frac{1}{2})}, \, 
\Phi_d \equiv 
\Phi_{\bf(10,2,2)}^{({\bf 1,2},-\frac{1}{2})}.
\ee
In the basis 
$\left\{ H_u^{10}, \overline{\Delta}_u, \Delta_u, \Phi_u \right\}$, 
the mass matrix is written as  
\be
\label{Mdoublet}
M_{\mathsf{doublet}} \equiv 
\left(
\begin{array}{cccc}
\begin{array}{c}
2 m_3 \\
\frac{\lambda_4 \phi_2}{\sqrt{10}} 
-\frac{\lambda_4 \phi_3}{2 \sqrt{5}} \\
-\frac{\lambda_3 \phi_2}{\sqrt{10}} 
-\frac{\lambda_3 \phi_3}{2 \sqrt{5}} \\
\frac{\lambda_3 v_R}{\sqrt{5}}
\end{array}
&\begin{array}{c}
\frac{\lambda_3 \phi_2}{\sqrt{10}} 
-\frac{\lambda_3 \phi_3}{2 \sqrt{5}} \\
m_2 +\frac{\lambda_2 \phi_2}{15 \sqrt{2}} 
-\frac{\lambda_2 \phi_3}{30} \\
0 \\
0
\end{array}
&\begin{array}{c}
-\frac{\lambda_4 \phi_2}{\sqrt{10}} 
-\frac{\lambda_4 \phi_3}{2 \sqrt{5}} \\
0 \\
m_2 +\frac{\lambda_2 \phi_2}{15 \sqrt{2}} 
+\frac{\lambda_2 \phi_3}{30}\\
-\frac{\lambda_2 v_R}{10}
\end{array}
&\begin{array}{c}
\frac{\lambda_4 \overline{v_R}}{\sqrt{5}} \\
0 \\
-\frac{\lambda_2 \overline{v_R}}{10} \\
2 m_1 + \frac{\lambda_1 \phi_2}{\sqrt{2}} 
+ \frac{\lambda_1 \phi_3}{2}
\end{array}
\end{array}
\right).  
\ee
The corresponding mass terms of the superpotential read 
\be
W_m = \left(H_u^{10}, \overline{\Delta}_u, \Delta_u, \Phi_u \right) 
\, M_{\mathsf{doublet}} \,
\left(H_d^{10}, \Delta_d, \overline{\Delta}_d, \Phi_d \right)^{\mathsf{T}}.  
\label{Wm}
\ee
The requirement of the existence of a zero mode leads to the 
following condition.  
\be 
\det{M_{\mathsf{doublet}}} = 0.  
\label{splittings}
\ee
For instance, in case of $\lambda_3 = 0$, 
$m_2 + \frac{\lambda_2 \phi_2}{15 \sqrt{2}} 
- \frac{\lambda_2 \phi_3}{30}=0$, 
we obtain a special solution to Eq. (\ref{splittings}), 
while it keeps a desirable vacuum and it does not produce 
any additional massless fields.  However, we proceed our 
arguments hereafter without using this special solution.  

We can diagonalize the mass matrix, $M_{\mathsf{doublet}}$ 
by a bi-unitary transformation.   
\be
U^{\ast} \,M_{\mathsf{doublet}} \,V^{\dagger}
= {\mathrm{diag}}(0, M_1, M_2, M_3).    
\ee
Then the mass eigenstates are written as  

\bea
\left(H_u, 
\, {\mathsf{h}}_u^1, \, {\mathsf{h}}_u^2, \, {\mathsf{h}}_u^3 \right)
&=& 
\left(H_u^{10}, {\overline{\Delta}}_u, \Delta_u, 
\Phi_u \right) \, U^{\mathsf{T}}, 
\nonumber\\
\left(H_d, 
\, {\mathsf{h}}_d^1, \, {\mathsf{h}}_d^2, \, {\mathsf{h}}_d^3 \right)
&=&
\left(H_d^{10}, \Delta_d, {\overline{\Delta}}_d, 
\Phi_d \right) \, V^{\mathsf{T}}. 
\label{UV}
\eea
The representations ${\bf 45}$ and/or ${\bf 54}$,
and higher dimensional operators, are not included in our minimal
model.  Therefore, we must set the "Doublet-Triplet
splittings" by hand as Eq. (\ref{splittings}).  
In the case of $\lambda_3 = \lambda_4$,  
Eq. (\ref{Mdoublet}) becomes symmetric, and 
$H_u$ and $H_d$ have the same coefficients in Eq. (\ref{UV}).  
This can not be accepted since it leads to the formal singularity 
in the low-energy Yukawa couplings (A matrix in Eq. (\ref{A})).  
Namely, it leads to the equality $Y_{u} = Y_{d}$, and therefore only 
the ratio of $Y_{10}$ and $Y_{126}$ can be determined from Eq. (\ref{A}).  
So we set $\lambda_3 \neq \lambda_4$ hereafter.  

By making the inverse transformation of Eq. (\ref{UV}), 
the following expressions are obtained,  
\be
H_u^{10} = \alpha_u H_u + \cdots, \,\,
H_d^{10} = \alpha_d H_d + \cdots, \,\,
\overline{\Delta}_u = \beta_u H_u +\cdots, \,\, 
\overline{\Delta}_d = \beta_d H_d +\cdots,
\label{mix}
\ee
where "$+\cdots$" represent the heavy Higgs fields, 
${\mathsf{h}}_{u, d}^i$ $(i=1,2,3)$ which are integrated out 
when considering the low-energy effective superpotential.  

Precisely, we can read off from Eq. (\ref{UV}) as      
\begin{eqnarray}
\alpha_u &=& (U^{\ast})_{11},\,\,
\beta_u = (U^{\ast})_{12},\,\,
\alpha_d = (V^{\ast})_{11},\,\,
\beta_d = (V^{\ast})_{13}.
\end{eqnarray}
Using the two pairs of the Higgs doublets, 
$H_{u, d}^{10}$ and $\overline{\Delta}_{u, d}$,
the Yukawa couplings of Eq. (\ref{yukawa}) are rewritten as
\begin{eqnarray}
W_Y &=& 
u_i^c 
\left(Y_{10}^{ij}\, H_u^{10} + Y_{126}^{ij}\, \overline{\Delta}_u 
\right) q_j 
+
d_i^c 
\left(Y_{10}^{ij}\, H_d^{10} + Y_{126}^{ij}\, \overline{\Delta}_d     
\right) q_j  
\nonumber\\ 
&+&
\nu_i^c 
\left(Y_{10}^{ij}\, H_u^{10} - 3 Y_{126}^{ij}\, \overline{\Delta}_u 
\right) \ell_j 
+
e_i^c 
\left(Y_{10}^{ij}\, H_d^{10}  - 3 Y_{126}^{ij}\, \overline{\Delta}_d  
\right) \ell_j   \nonumber \\
&+&
\nu_i^c 
\left( Y_{126}^{ij} v_R \right) \nu_j^c .   
\end{eqnarray}
By using Eq. (\ref{mix}), we obtain 
the low-energy effective superpotential which 
is described by only the light Higgs doublets 
$H_u$ and $H_d$,  
\begin{eqnarray}
W_{\mathrm{eff}} 
&=& 
u^c_i \left(\alpha_u Y_{10}^{ij} + \beta_u Y_{126}^{ij} \right) 
H_u \, q_j 
+ 
d^c_i \left(\alpha_d Y_{10}^{ij} + \beta_d Y_{126}^{ij} \right) 
H_d \, q_j 
\nonumber\\ 
&+& 
\nu^c_i \left(\alpha_u Y_{10}^{ij} -3 \beta_u Y_{126}^{ij} \right)
H_u \, \ell_j 
+
e^c_i \left(\alpha_d Y_{10}^{ij} -3 \beta_d Y_{126}^{ij} \right)
H_d \, \ell_j 
\nonumber\\ 
&+& 
\nu^c_i \left( Y_{126}^{ij} v_R \right) \nu^c_j
\nonumber\\
&+& \mu_{\mathrm{eff}} \, H_u H_d.
\label{Yukawa3}
\end{eqnarray} 
Here we have assumed that some mechanism, like the Giudice-Masiero 
mechanism {\cite{GM}} in supergravity, may produce the effective $\mu$ term, 
$\mu_{\mathrm{eff}}$ for the light Higgs doublets.  
\section{Proton decay}
After the symmetry breaking from SO(10) to 
$SU(3)_C \times SU(2)_L \times U(1)_Y$, 
the generic Yukawa interactions between the matter fields 
and the color triplet Higgs fields are given by  
\bea
W_{Y}&=&
Y_{10}^{ij}\,H_{\overline{T}}
\left(q_i \ell_j + u^c_i d^c_j \right)
+Y_{126}^{ij}\,\overline{\Delta}_{\overline{T}}
\left(q_i \ell_j + u^c_i d^c_j \right)
\nonumber\\
&+&
Y_{10}^{ij}\,H_T 
\left(\frac{1}{2} q_i q_j 
+u^c_i e^c_j 
+d^c_i \nu^c_j \right)
\nonumber\\
&+&Y_{126}^{ij}\,\overline{\Delta}_T
\left(\frac{1}{2} q_i q_j 
+u^c_i e^c_j + d^c_i \nu^c_{j} 
\right)
\nonumber\\
&+&
Y_{126}^{ij}\,{\overline{\Delta}}_T^{\prime}
\left(u^c_i e^c_j + d^c_i \nu^c_{j} 
\right).
\eea
Here we have defined  
\be
H_{\overline{T}} 
\equiv
H^{{\bf(\overline{3},1},\frac{1}{3})}_{\bf(6,1,1)}\,,\,\,
H_T 
\equiv
H^{{\bf(3,1},-\frac{1}{3})}_{\bf(6,1,1)}\,,\,\,
\overline{\Delta}_{\overline{T}} 
\equiv
\overline{\Delta}^{{\bf(\overline{3},1},\frac{1}{3})}_{\bf(6,1,1)}\,,\,\,
\overline{\Delta}_T 
\equiv
\overline{\Delta}^{{\bf(3,1},-\frac{1}{3})}_{\bf(6,1,1)}\,,\,\,
{\overline{\Delta}}_T^{\prime} 
\equiv
\overline{\Delta}^{{\bf(3,1},-\frac{1}{3})}_{\bf(10,1,3)}\,. 
\ee
For later use we define
\be
{\Delta}_{\overline{T}} 
\equiv
\Delta^{{\bf(\overline{3},1},\frac{1}{3})}_{\bf(6,1,1)}\,,\,\, 
{\Delta}_{T} 
\equiv
\Delta^{{\bf(3,1},-\frac{1}{3})}_{\bf(6,1,1)}\,,\,\, 
{\Delta}_{\overline{T}}^{\prime} 
\equiv
\Delta^{{\bf(\overline{3},1},\frac{1}{3})}_{\bf(\overline{10},1,3)}\,,\,\,
\Phi_{\overline T} 
\equiv
\Phi^{{\bf(\overline{3},1},\frac{1}{3})}_{\bf(15,1,3)}\,,\,\,
\Phi_T 
\equiv
\Phi^{{\bf(3,1},-\frac{1}{3})}_{\bf(15,1,3)}\,. 
\ee
In the basis $
\left\{
H_{\overline T}, \Delta_{\overline T}, 
{\overline \Delta}_{\overline T}, 
\Phi_{\overline T}, 
\Delta_{\overline T}^{\prime} \right\}$, 
the mass matrix reads
\be
M_{\mathsf{triplet}} \equiv
\left(
\begin{array}{ccccc}
\begin{array}{c}
2 m_3 \\
-\frac{\lambda_3 \phi_1}{\sqrt{10}} - \frac{\lambda_3 \phi_2}{\sqrt{30}} \\
-\frac{\lambda_4 \phi_1}{\sqrt{10}} + \frac{\lambda_4 \phi_2}{\sqrt{30}} \\
\frac{\lambda_3 v_R}{\sqrt{5}} \\
-\frac{\sqrt{2}\lambda_3 \phi_3}{\sqrt{15}}
\end{array}
&\begin{array}{c}
-\frac{\lambda_4 \phi_1}{\sqrt{10}} - \frac{\lambda_4 \phi_2}{\sqrt{30}} \\
m_2 \\
0 \\
-\frac{\lambda_2 v_R}{10 \sqrt{3}} \\
\frac{\lambda_2 \phi_3}{15 \sqrt{2}} 
\end{array}
&\begin{array}{c}
-\frac{\lambda_3 \phi_1}{\sqrt{10}} + \frac{\lambda_3 \phi_2}{\sqrt{30}} \\
0 \\
m_2 \\
0 \\
0
\end{array}
&\begin{array}{c}
\frac{\lambda_4 \overline{v_R}}{\sqrt{5}} \\
-\frac{\lambda_2 \overline{v_R}}{10\sqrt{3}} \\
0 \\
\overline{m}_{44}  \\
-\frac{\lambda_2 \overline{v_R}}{5 \sqrt{6}}
\end{array}
&\begin{array}{c}
-\frac{\sqrt{2}\lambda_4 \phi_3}{\sqrt{15}} \\
\frac{\lambda_2 \phi_3}{15 \sqrt{2}} \\
0 \\
-\frac{\lambda_2 v_R}{5 \sqrt{6}} \\
\overline{m}_{55}
\end{array}
\end{array}
\right),  
\ee
where $\overline{m}_{44} 
\equiv 
2 m_1 + \frac{\lambda_1 \phi_1}{\sqrt{6}}
+ \frac{\lambda_1 \phi_2 }{3 \sqrt{2}}
+ \frac{2 \lambda_1 \phi_3}{3}$ 
and 
$\overline{m}_{55} 
\equiv 
m_2 +\frac{\lambda_2 \phi_1}{10 \sqrt{6}}
+\frac{\lambda_2 \phi_2}{30 \sqrt{2}}$.  

The corresponding mass terms of the superpotential read 
\be
W_m = \left(
H_{\overline T}, \Delta_{\overline T}, 
{\overline \Delta}_{\overline T}, \Phi_{\overline T}, 
\Delta_{\overline T}^{\prime}
\right) \,
M_{\mathsf{triplet}} \,
\left(
H_T, {\overline \Delta}_T, 
\Delta_T, \Phi_T, 
{\overline{\Delta}}_T^{\prime}
\right)^{\mathsf{T}}.  
\label{WmT}
\ee
Here we integrate out the color triplet Higgs fields,  
$\Delta_T$ and $\Phi_T$,  
which do not appear in the Yukawa interaction 
with matter, Eq. (\ref{yukawa}),  
\be
\left[
\begin{array}{c}
\overline{\Delta}_{\overline{T}} \\
\Phi_{\overline{T}} 
\end{array}
\right]
=-\frac{1}{\cal D} \,
\left[
\begin{array}{c}
\left(2 m_1 + \frac{\lambda_1 \phi_1}{\sqrt{6}}
+ \frac{\lambda_1 \phi_2 }{3 \sqrt{2}}
+ \frac{2 \lambda_1 \phi_3}{3} \right) \cdot
\left(-\frac{\lambda_3\phi_1}{\sqrt{10}} 
+ \frac{\lambda_3\phi_2}{\sqrt{30}} \right) \, 
H_{\overline{T}} 
\\
m_2 \cdot
\frac{\lambda_4 \overline{v_R}}{\sqrt{5}} \,
H_{\overline{T}}
- m_2 \cdot \frac{\lambda_2 \overline{v_R}}{10 \sqrt{3}} \,
\Delta_{\overline{T}} 
- m_2 \cdot \frac{\lambda_2 \overline{v_R}}{5 \sqrt{6}} \,
\Delta_{\overline{T}}^{\prime}
\end{array}
\right].
\ee
where ${\cal D} \equiv m_2 \cdot
\left(2 m_1 + \frac{\lambda_1 \phi_1}{\sqrt{6}}
+ \frac{\lambda_1 \phi_2 }{3 \sqrt{2}}
+ \frac{2 \lambda_1 \phi_3}{3} \right) $.  

Putting this into the original mass terms of the superpotential 
Eq. (\ref{WmT}), we can obtain the following mass terms for the 
color triplet Higgs fields, 
\bea
W_{m}^{\mathrm{eff}} &=&
\left(
H_{\overline T}, \Delta_{\overline T}, 
{\Delta}_{\overline T}^{\prime}
\right) \,
M_{\mathsf{triplet}}^{\mathrm{eff}} \,
\left(
H_T, {\overline \Delta}_T, {\overline{\Delta}}_T^{\prime}
\right)^{\mathsf{T}}.  
\label{WmTprime}
\eea
Here the explicit forms of the elements of this mass matrix, 
$M_{\mathsf{triplet}}^{\mathrm{eff}} =\left\{m_{ij} \right\}$ are given as follows, 

\bea
m_{11} &\equiv& 2 m_3 
- \frac{1}{\cal D} \,
\left[
\left(-\frac{\lambda_4 \phi_1}{\sqrt{10}}
+\frac{\lambda_4 \phi_2}{\sqrt{30}} \right) \cdot
\left(2 m_1 + \frac{\lambda_1 \phi_1}{\sqrt{6}}
+ \frac{\lambda_1 \phi_2 }{3 \sqrt{2}}
+ \frac{2 \lambda_1 \phi_3}{3} \right) \cdot
\left(-\frac{\lambda_3 \phi_1}{\sqrt{10}} 
+\frac{\lambda_3 \phi_2}{\sqrt{30}} \right) 
\right.
\nonumber\\
&+& \left. 
\frac{\lambda_3 v_R}{\sqrt{5}} \cdot
m_2 \cdot
\frac{\lambda_4 \overline{v_R}}{\sqrt{5}}
\right],
\nonumber\\
m_{12} &\equiv& -\frac{\lambda_4 \phi_1}{\sqrt{10}} 
-\frac{\lambda_4 \phi_2}{\sqrt{30}}
+ \frac{1}{\cal D} \,
\frac{\lambda_2 v_R}{10 \sqrt{3}} \cdot
m_2 \cdot
\frac{\lambda_4 \overline{v_R}}{\sqrt{5}},
\nonumber\\
m_{13} &\equiv& 
-\frac{\sqrt{2} \lambda_4 \phi_3}{\sqrt{15}} 
+\frac{1}{\cal D} \, 
\frac{\lambda_2 v_R}{5 \sqrt{6}} \cdot
m_2 \cdot
\frac{\lambda_4 \overline{v_R}}{\sqrt{5}}, 
\nonumber\\
m_{21} &\equiv& 
-\frac{\lambda_3 \phi_1}{\sqrt{10}} - \frac{\lambda_3 \phi_2}{\sqrt{30}}
+ \frac{1}{\cal D} \,
\frac{\lambda_3 v_R}{\sqrt{5}} \cdot
m_2 \cdot
\frac{\lambda_2 \overline{v_R}}{10 \sqrt{3}},
\nonumber\\
m_{22} &\equiv& m_2
- \frac{1}{\cal D} \,
\frac{\lambda_2 v_R}{10 \sqrt{3}} \cdot
m_2 \cdot
\frac{\lambda_2 \overline{v_R}}{10 \sqrt{3}},
\nonumber\\
m_{23} &\equiv& 
\frac{\lambda_2 \phi_3}{15 \sqrt{2}}
- \frac{1}{\cal D} \,
\frac{\lambda_2 v_R}{10 \sqrt{3}} \cdot
m_2 \cdot
\frac{\lambda_2 \overline{v_R}}{10 \sqrt{3}},
\nonumber\\
m_{31} &\equiv& -\frac{\sqrt{2} \lambda_3 \phi_3}{\sqrt{15}} 
+ \frac{1}{\cal D} \,
\frac{\lambda_3 v_R}{\sqrt{5}} \cdot
m_2 \cdot
\frac{\lambda_2 \overline{v_R}}{5 \sqrt{6}},  
\nonumber\\
m_{32} &\equiv& 
\frac{\lambda_2 \phi_3}{15 \sqrt{2}}
- \frac{1}{\cal D} \,
\frac{\lambda_2 v_R}{10 \sqrt{3}} \cdot
m_2 \cdot
\frac{\lambda_2 \overline{v_R}}{5 \sqrt{6}},
\nonumber\\
m_{33} &\equiv& 
m_2 +\frac{\lambda_2 \phi_1}{10 \sqrt{6}}
+\frac{\lambda_2 \phi_2}{30 \sqrt{2}} 
- \frac{1}{\cal D} \,
\frac{\lambda_2 v_R}{5 \sqrt{6}} \cdot
m_2 \cdot 
\frac{\lambda_2 \overline{v_R}}{5 \sqrt{6}}.
\eea
Moreover, integrating out the color triplet Higgs field 
${\Delta}_{\overline{T}}^{\prime}$, 
we obtain the effective Yukawa interactions between 
the matter fields and the color triplet Higgs fields as  
\bea
W_{Y} &=& 
Y_{10}^{ij} \, 
H_{\overline{T}} \, 
\left(q_i \ell_j + u^c_i d^c_j \right) 
+ Y_{126}^{ij} \, 
\overline{\Delta}_{\overline{T}} \, 
\left(q_i \ell_j + u^c_i d^c_j \right)
\nonumber\\
&+&
Y_{10}^{ij} \, 
H_T \,
\frac{1}{2} q_i q_j 
\nonumber\\
&+&
\left(Y_{10}^{ij} - \frac{m_{31}}{m_{33}} \, Y_{126}^{ij} \right) \, 
H_T \, 
\left( u^c_i e^c_j + d^c_i \nu^c_j \right)
\nonumber\\
&+&Y_{126}^{ij} \, 
\overline{\Delta}_T \,
\frac{1}{2} q_i q_j 
\nonumber\\
&+& \left(1 -\frac{m_{32}}{m_{33}} \right) 
Y_{126}^{ij} \, 
\overline{\Delta}_T \,
\left(u^c_i e^c_j + d^c_i \nu^c_j \right). 
\label{yukawa2}
\eea
Then the effective mass terms for 
the remaining color triplet Higgs fields are written as  
\bea
W_{m}^{\mathrm{eff}}&=&
H_{\overline{T}}\left(a\, H_T +b \, \overline{\Delta}_T \right)
\nonumber\\
&+&
\overline{\Delta}_{\overline{T}}\left(c \,H_T +d\, \overline{\Delta}_T \right)
\nonumber\\
&\equiv&
\left(H_{\overline{T}},\,\overline{\Delta}_{\overline{T}} \right)
\,M_T \,
\left(
\begin{array}{c}
H_T \\
\overline{\Delta}_T 
\end{array}
\right),
\label{mass}
\eea
where $a$, $b$, $c$, $d$ are defined by 
\bea
a &\equiv& m_{11} -\frac{m_{13}}{m_{33}} \cdot m_{31}, \quad
b \equiv m_{12} -\frac{m_{13}}{m_{33}} \cdot m_{32}, 
\nonumber\\
c &\equiv& m_{21} -\frac{m_{23}}{m_{33}} \cdot m_{31}, \quad
d \equiv m_{22} -\frac{m_{23}}{m_{33}} \cdot m_{32}.   
\eea
Combining the Eqs. (\ref{yukawa2}) and (\ref{mass}) leads to 
the effective dimension-five interactions after integrating out 
the remaining color triplet Higgs fields \cite{Sakai},  
\be
-W_5 =  C_L^{ijkl} \, \frac{1}{2} q_i q_j q_k \ell_l
+ C_R^{ijkl} \, u^c_i e^c_j u^c_k d^c_l ,
\label{dim5}
\ee
inducing the dangerous proton decay.  
Here, $C_L$ and $C_R$ are given by the Yukawa coupling 
matrices at the GUT scale, $M_G$    
\bea
C_L^{ijkl} (M_{G})
&=&
\left(Y_{10}^{ij},\, Y_{126}^{ij} \right)
M_T^{-1}
\left(
\begin{array}{c}
Y_{10}^{kl} \\
Y_{126}^{kl}
\end{array} 
\right),
\nonumber\\
C_R^{ijkl} (M_{G})
&=&
\left( Y_{10}^{ij} - \frac{m_{13}}{m_{33}} \,Y_{126}^{ij}, \, 
\left(1 -\frac{m_{32}}{m_{33}} \right) \,Y_{126}^{ij} \right)
M_T^{-1}
\left(
\begin{array}{c}
Y_{10}^{kl} \\
Y_{126}^{kl}
\end{array} 
\right).
\eea
Note that
\bea
\left(
\begin{array}{c}
Y_{10} \\
Y_{126}
\end{array}
\right)
&=&
\left(
\begin{array}{cc}
\begin{array}{c}
\alpha_u \quad \\
\alpha_d \quad
\end{array}
\begin{array}{c}
\beta_u \\
\beta_d
\end{array}
\end{array}
\right)^{-1}
\left(
\begin{array}{c}
Y_{u} \\
Y_{d}
\end{array}
\right)
\nonumber\\
&\equiv&
A^{-1} \,\left(
\begin{array}{c}
Y_{u} \\
Y_{d}
\end{array}
\right).  
\label{A}
\eea
Thus we have 
\bea
C_L^{ijkl} 
&=& 
\left(Y_{u}^{ij},~Y_{d}^{ij} \right)
\left(
A\,
M_T
\,
A^{\mathsf{T}}
\right)^{-1}
\left(
\begin{array}{c}
Y_{u}^{kl} \\
Y_{d}^{kl}
\end{array} 
\right).
\eea
We make use of this expressions in order to evaluate 
the renormalization group effects on the 
Wilson coefficients $C_{L}^{ijkl}$ and $C_{R}^{ijkl}$.  
Without loss of generality, we can use the basis 
where $Y_u$ is real and diagonal, 
\be
Y_u = \frac{1}{v \sin\beta}\, {\mathrm{diag}}(m_u, m_c, m_t), 
\ee
with $v \simeq 174.1$ [GeV].  
Since $Y_d$ is a symmetric matrix, 
it can be described as 
\be
Y_d =\frac{1}{v \cos\beta}\, {\overline{V}}_{\mathrm{CKM}}^{\,\ast} \,
{\mathrm{diag}}(m_d, m_s, m_b) \, 
{\overline{V}}_{\mathrm{CKM}}^\dagger,
\ee
by using a unitary matrix 
\be
{\overline{V}}_{\mathrm{CKM}} 
\equiv e^{i \alpha_1} \,e^{i \alpha_2 \lambda_3} \,e^{i \alpha_3 \lambda_8}\, 
  V_{\mathrm{CKM}} \,
  e^{i \beta_2 \lambda_3} \,e^{i \beta_3 \lambda_8}, 
\ee
where $\lambda_3$, $\lambda_8$ are the Gell-Mann matrices and 
$V_{\mathrm{CKM}}$ is 
the Cabibbo-Kobayashi-Maskawa (CKM) mixing matrix \cite{KM}.  
\footnote{In Ref. \cite{fukuyama}, we set these phases $\alpha_i$ 
$(i = 1,2,3)$, $\beta_i$ $(i = 2,3)$ to zero or $\pi$.  }

The complete anti-symmetry in the color indices 
requires that the dimension-five operator Eq. (\ref{dim5}) 
possesses the flavor non-diagonal indices \cite{Dimopoulos}.  
As a consequence, the dominant decay mode is 
$p \rightarrow K^{+} \bar{\nu}$.  This fact implies 
that the chargino dressing diagrams dominate 
over the gluino and the neutralino dressing 
diagrams \cite{Nath}.  

In the components form, the dimension-five operators 
at the SUSY breaking scale, $M_{\mathrm{SUSY}}$ are written as 
\bea
{\mathcal L}_5 &=& 
C_{L}^{(\tilde{u} \tilde{d} u e) XYij}
\widetilde{u_{X}} \widetilde{d_{Y}} u_{Li} e_{Lj}
+C_{L}^{(\tilde{u} \tilde{u} d e) XYij}
\frac{1}{2}
\widetilde{u_{X}} \widetilde{u_{Y}} d_{Li} e_{Lj}
\nonumber\\
&+&
C_{R}^{(\tilde{u} \tilde{d} u e) XYij}
\widetilde{u_{X}} \widetilde{d_{Y}} u_{Ri} e_{Rj}
+C_{R}^{(\tilde{u} \tilde{u} d e) XYij}
\frac{1}{2}
\widetilde{u_{X}} \widetilde{u_{Y}} d_{Ri} e_{Rj}
\nonumber\\
&+&
C_{L}^{(\tilde{u} \tilde{d} d \nu) XYij}
\widetilde{u_{X}} \widetilde{d_{Y}} d_{Li} \nu_{Lj}
+C_{L}^{(\tilde{d} \tilde{d} u \nu) XYij}
\frac{1}{2}
\widetilde{d_{X}} \widetilde{d_{Y}} u_{Li} \nu_{Lj}
\nonumber\\
&+&
C_{L}^{(\tilde{u} \tilde{e} u d) XYij}
\widetilde{u_{X}} \widetilde{e_{Y}} u_{Li} d_{Lj}
+C_{L}^{(\tilde{d} \tilde{e} u u) XYij}
\frac{1}{2}
\widetilde{d_{X}} \widetilde{e_{Y}} u_{Li} u_{Lj}
\nonumber\\
&+&
C_{R}^{(\tilde{u} \tilde{e} u d) XYij}
\widetilde{u_{X}} \widetilde{e_{Y}} u_{Ri} d_{Rj}
+C_{R}^{(\tilde{d} \tilde{e} u u) XYij}
\frac{1}{2}
\widetilde{d_{X}} \widetilde{e_{Y}} u_{Ri} u_{Rj}
\nonumber\\
&+&
C_{L}^{(\tilde{d} \tilde{\nu} u d) XYij}
\widetilde{d_{X}} \widetilde{\nu_{Y}} u_{Li} d_{Lj}
+C_{L}^{(\tilde{u} \tilde{\nu} d d) XYij}
\frac{1}{2}
\widetilde{u_{X}} \widetilde{\nu_{Y}} d_{Li} d_{Lj}.
\eea
The coefficients are obtained from the coefficients of the original dimension-five 
operators including their renormalizion 
from $M_{G}$ to $M_{\mathrm{SUSY}}$.  Their explicit forms are 
found in Appendix A.  
After the sparticles dressing, we obtain the following type of 
dimension-six operators causing nucleon decays,
\bea
{\mathcal L}_6&=&
\frac{1}{16 \pi^2}\left[
C_{LL}^{(udue)ij} (u_{L}d_{Li})(u_{L}e_{Lj})
+C_{RL}^{(udue)ij} (u_{R}d_{Ri})(u_{L}e_{Lj})
\right.
\nonumber\\
&+&
C_{LR}^{(udue)ij} (u_{L}d_{Li})(u_{R}e_{Rj})
+C_{RR}^{(udue)ij} (u_{R}d_{Ri})(u_{R}e_{Rj})
\nonumber\\
&+&
C_{LL}^{(udd\nu)ijk} (u_{L}d_{Li})(d_{Lj}\nu_{Lk})
+C_{RL}^{(udd\nu)ijk} (u_{R}d_{Ri})(d_{Lj}\nu_{Lk})
\nonumber\\
&+&
\left.
C_{RL}^{(ddu\nu)ijk} \frac{1}{2} (d_{Ri}d_{Rj})(u_{L}\nu_{Lk})
\right].
\eea
These operators should be renormalized from $M_{\mathrm{SUSY}}$ 
to $M_{Z}$ and further to the hadronization scale 
($\mu_{{\mathrm had}}$) $\approx$1  [GeV].  Then 
the effective four-Fermi Lagrangian is converted to 
a hadronic Lagrangian by using the chiral Lagrangian method 
\cite{Claudson}\cite{Chadha}.  Details are given 
in Appendices B and C.  

For the decay mode $p \rightarrow K^{+} \bar{\nu_i}$, 
the partial decay rate is given by the formula
\be
\Gamma(p \rightarrow K^{+} \bar{\nu_i})
=\frac{m_p}{32 \pi}\left(1- \frac{m_{{K}^{+}}^2}{m_p^2} \right)^2
\frac{1}{{f_{\pi}}^2}\, |{\mathcal A}(p \rightarrow K^{+} \bar{\nu_i})|^2.
\ee
Here $m_p=0.938\,[{\mathrm GeV}]$ is the proton mass, 
$m_{K^{+}}=0.493\,[{\mathrm GeV}]$ is the kaon mass 
and $f_{\pi}=0.131\,[{\mathrm GeV}]$ is the pion decay constant.  

The amplitude ${\mathcal A}(p \rightarrow K^{+} \bar{\nu_i})$ 
for $p \rightarrow K^{+} \bar{\nu_i}$ reads \cite{Goto}
\bea
{\mathcal A}(p \rightarrow K^{+} \bar{\nu_i})
&=&\left[ \beta C_{LL}^{(udd\nu)21i} 
+ \alpha C_{RL}^{(udd\nu)21i} \right] 
\frac{2 m_p}{3 m_B} D 
\nonumber\\
&+&\left[ \beta C_{LL}^{(udd\nu)12i} 
+ \alpha C_{RL}^{(udd\nu)12i} \right] 
\left[1+\frac{m_p}{3 m_B}\left(3 F+D \right) \right] 
\nonumber\\
&+&
\alpha C_{RL}^{(ddu\nu)12i} 
\left[1-\frac{m_p}{3 m_B}\left(3 F-D \right) \right]. 
\eea
Here $m_{B}=1.150 \,[{\mathrm GeV}]$ is an averaged baryon mass, 
$F=0.44$, $D=0.81$ are the parameters in terms of which 
the octet-baryon axial-vector form factors are expressed
and $\alpha$, $\beta$ are the hadron matrix elements 
which are defined by \cite{Brodsky}
\bea
\alpha u_L (\bf k) &=& \langle 0 | 
d_R u_R u_L |p (\bf k) \rangle,
\nonumber\\
\beta u_L (\bf k) &=& \langle 0 | 
d_L u_L u_R |p (\bf k) \rangle.
\eea
$u_L (\bf k)$ denote the left-handed components of the 
proton wave function. It is known that 
$|\alpha|=|\beta|$, and $\beta$ is in the range \cite{Brodsky}
\be
0.003 \,[{\mathrm GeV}^3] \leq \beta \leq 0.03 \,[{\mathrm GeV}^3].
\ee
From the recent lattice calculations, one group reported that \cite{JLQCD}  
\bea
\alpha &=& -( 0.015 \pm 0.001 ) \,[{\mathrm GeV}^3],
\nonumber\\
\beta &=&  0.014 \pm 0.001  \,[{\mathrm GeV}^3].
\eea
But the other group reported the smaller values \cite{RBC}
\bea
\alpha &=& -( 0.006 \pm 0.001 ) \,[{\mathrm GeV}^3],
\nonumber\\
\beta &=&  0.007 \pm 0.001  \,[{\mathrm GeV}^3].
\eea
\section{Gauge couplings unification}
In general, the gauge couplings unification imposes constraints 
on the mass spectrum of many varieties of Higgs fields \cite{lucas}.  
Our strategy is a generic one that all of the dimensionless coefficients 
should remain of order one to preserve the perturbative limit 
and put all the VEVs at the GUT scale in order to realize 
the simple gauge couplings unification picture.    
For the numerical evaluation, 
we use the one-loop renormalization group equations (RGEs) 
in the $\overline{{\mathrm DR}}$ scheme \cite{Siegel}, 
\footnote{$\overline{{\mathrm DR}}$ uses dimensional regularization 
through dimensional reduction with modified minimal subtraction.  }
\footnote{Here we assume, for simplicity, all the mass eigenvalues of 
the Higgs fields are smaller than $M_G$ and 
all the masses of the gauge fields lie around $M_G$.  
In the other cases, the formula becomes quite complicated.  }  
\be
\frac{1}{\alpha_i \left(M_G \right)}
=
\frac{1}{\alpha_i \left(M_Z \right)}|_{\overline{\mathrm MS}}
-\frac{C_2\left(G_i \right)}{12 \pi}
+\frac{1}{2 \pi}\left[
b_i\log{\left(\frac{M_Z}{M_G}\right)}
+\sum_{\zeta} {b_i^{\zeta}
\log{\left(\frac{\det^{\prime}{M_\zeta}}{M_G^{{\mathrm rank}(M_\zeta )}}
\right)}}
\right],  
\label{RGE}
\ee
where $C_2$ is the quadratic Casimir operator; $C_2 \left(SU(3) \right)=3$, 
$C_2 \left(SU(2) \right)=2$, $C_2 \left(U(1) \right)=0$, and 
$\zeta$ denotes the Higgs fields which have the corresponding 
gauge quantum numbers.  $M_{\zeta}$ is it's mass matrix and  
"$\det^{\prime}$ " means that the determinant should be taken 
excluding  the zero modes.  
$b_i$ and $b_i^\zeta$ are the $\beta$ function coefficients ; 
$b_3 =-3$, $b_2 = 1$, $b_1 = \frac{33}{5}$, and $b_i^\zeta$ 
are given in Tables \ref{tab1} and \ref{tab2}.  For $\alpha_i 
\left(M_Z \right)|_{\overline{\mathrm MS}}$, we use the following values.  
\bea
\alpha_3 \left(M_Z \right)|_{\overline{\mathrm MS}}
&=& \alpha_s \left(M_Z \right),
\\
\alpha_2 \left(M_Z \right)|_{\overline{\mathrm MS}}
&=& 
\alpha \left(M_Z \right)/\sin^2 {\theta_{\mathrm W}}\left(M_Z \right),
\\
\alpha_1 \left(M_Z \right)|_{\overline{\mathrm MS}}
&=&
\frac{5}{3} \,
\alpha \left(M_Z \right)/
\left(1 - \sin^2 {\theta_{\mathrm W}}\left(M_Z \right) \right),
\eea
with \cite{PDG}
\be
\alpha_s \left(M_Z \right)=0.1172, \quad
\alpha \left(M_Z \right)=1/128.92, \quad
\sin^2 {\theta_{\mathrm W}} \left(M_Z \right) =0.23113.  
\ee
Excluding the fields which mix with the would-be NG 
fields and the fields with $SU(3)_C \times SU(2)_L 
\times U(1)_Y$ quantum numbers, 
$[({\bf1,2},\frac{1}{2})+{\mathrm h.c.}]$ ("Higgs doublet") and 
$[({\bf \overline{3},1},\frac{1}{3}) +{\mathrm h.c.}]$
(color triplet Higgs fields), the massive fields are given 
as follows.  

For ${\bf \overline{126}}$ and ${\bf 126}$ representation 
fields, their quantum numbers, the masses and their $\beta$ function 
coefficients are given in Table \ref{tab1}.  

For {\bf 210} representation field, 
their quantum numbers, the masses and their $\beta$ function 
coefficients are given in Table \ref{tab2}.  
\begin{table}
\caption{The mass matrices and the $\beta$ function 
coefficients for ${\bf \overline{126}}$ and ${\bf 126}$.}
\label{tab1}
\begin{center}
\begin{tabular}{|c|c|c|c|c|}
quantum numbers
& mass matrices, or mass eigenvalues
& $b_3^{\zeta}$
& $b_2^{\zeta}$
& $b_1^{\zeta}$
\\
\hline
$\left({\bf 8,2},\frac{1}{2} \right)+{\mathrm h.c.}$
& 
$ 
\left(
\begin{array}{cc}
\begin{array}{c}
m_2 - \frac{\lambda_2 \phi_2}{30 \sqrt{2}} 
- \frac{\lambda_2 \phi_3}{60} \\
0 
\end{array}
&\begin{array}{c}
0 \\
m_2 - \frac{\lambda_2 \phi_2}{30 \sqrt{2}}
+ \frac{\lambda_2 \phi_3}{60}
\end{array}
\end{array}
\right)
$
& 12
& 8
& $\frac{24}{5}$
\\
\hline
$\left({\bf 6,3},\frac{1}{3} \right)+{\mathrm h.c.}$
& $ m_2 - \frac{\lambda_2 \phi_1}{10 \sqrt{6}} 
-\frac{\lambda_2 \phi_2}{30 \sqrt{2}} $
& 15
& 24
& $\frac{12}{5}$
\\
\hline
$\left({\bf 6,1},\frac{4}{3} \right)+{\mathrm h.c.}$
& 
$ m_2 + \frac{\lambda_2 \phi_1}{10 \sqrt{6}} 
- \frac{\lambda_2 \phi_2}{30 \sqrt{2}} 
- \frac{\lambda_2 \phi_3}{30} $
& 5
& 0
& $\frac{64}{5}$
\\
\hline
$\left({\bf \overline{6},1},\frac{2}{3} \right)+{\mathrm h.c.}$
&
$ m_2 + \frac{\lambda_2 \phi_1}{10 \sqrt{6}} 
-\frac{\lambda_2 \phi_2}{30 \sqrt{2}} 
+\frac{\lambda_2 \phi_3}{30} 
$
& 5
& 0
& $\frac{16}{5}$
\\
\hline
$\left({\bf 6,1},\frac{1}{3} \right)+{\mathrm h.c.}$
&
$ m_2 + \frac{\lambda_2 \phi_1}{10 \sqrt{6}} 
- \frac{\lambda_2 \phi_2}{30 \sqrt{2}} 
$
& 5
& 0
& $\frac{4}{5}$
\\
\hline
$\left({\bf \overline{3},3},\frac{1}{3} \right)+{\mathrm h.c.}$
&$ m_2 - \frac{\lambda_2 \phi_1}{10 \sqrt{6}} 
+ \frac{\lambda_2 \phi_2}{30 \sqrt{2}} 
$
& 3
& 12
& $\frac{6}{5}$
\\
\hline
$\left({\bf 3,2},\frac{7}{6} \right)+{\mathrm h.c.}$
&
$ 
\left(
\begin{array}{cc}
\begin{array}{c}
m_2 + \frac{\lambda_2 \phi_2}{30 \sqrt{2}} - \frac{\lambda_2 \phi_3}{60} \\
0 
\end{array}
&\begin{array}{c}
0 \\
m_2 + \frac{\lambda_2 \phi_2}{30 \sqrt{2}} - \frac{\lambda_2 \phi_3}{20}  
\end{array}
\end{array}
\right)
$
& 2
& 3
& $\frac{49}{5}$
\\
\hline
$\left({\bf \overline{3},1},\frac{4}{3} \right)+{\mathrm h.c.}$
& 
$ m_2 + \frac{\lambda_2 \phi_1}{10 \sqrt{6}} 
+ \frac{\lambda_2 \phi_2}{30 \sqrt{2}} 
- \frac{\lambda_2 \phi_3}{30}
$
& 1
& 0
& $\frac{32}{5}$
\\
\hline
$\left({\bf 1,3},1 \right)+{\mathrm h.c.}$
&
$ m_2 - \frac{\lambda_2 \phi_1}{10 \sqrt{6}} 
+ \frac{\lambda_2 \phi_2}{10 \sqrt{2}}
$
& 0
& 4
& $\frac{18}{5}$
\\
\hline
$\left({\bf 1,1},2 \right)+{\mathrm h.c.}$
&
$ m_2 + \frac{\lambda_2 \phi_1}{10 \sqrt{6}} 
+ \frac{\lambda_2 \phi_2}{10 \sqrt{2}} 
- \frac{\lambda_2 \phi_3}{10} 
$
& 0
& 0
& $\frac{24}{5}$
\\
\end{tabular} 
\end{center}
\end{table}
\begin{table}
\caption{The mass matrices and the $\beta$ function 
coefficients for ${\bf{210}}$.}
\label{tab2}
\begin{center}
\begin{tabular}{|c|c|c|c|c|}
quantum numbers
& mass matrices, or mass eigenvalues
& $b_3^{\zeta}$
& $b_2^{\zeta}$
& $b_1^{\zeta}$
\\
\hline
$\left({\bf 8,3},0 \right)$
& $ 2 m_1 - \frac{\lambda_1 \phi_1}{\sqrt{6}} 
- \frac{\lambda_1 \phi_2}{3 \sqrt{2}} 
$ 
& 9
& 16
& 0
\\
\hline
$\left({\bf 8,1},1 \right)+{\mathrm h.c.}$
& 
$ 2 m_1 + \frac{\lambda_1 \phi_1}{\sqrt{6}} 
- \frac{\lambda_1 \phi_2}{3 \sqrt{2}}  
$
& 6
& 0
& $\frac{48}{5}$
\\
\hline
$\left({\bf 8,1},0 \right)$
& 
$ \left(
\begin{array}{cc}
\begin{array}{c}
2 m_1 - \frac{\lambda_1 \phi_2}{3 \sqrt{2}} \\
\frac{\lambda_1 \phi_3}{3 \sqrt{2}} 
\end{array}
&\begin{array}{c}
\frac{\lambda_1 \phi_3}{3 \sqrt{2}} \\
2 m_1 + \frac{\lambda_1 \phi_1}{\sqrt{6}} 
- \frac{\lambda_1 \phi_2}{3 \sqrt{2}}
\end{array}
\end{array}
\right)
$
& 3
& 0
& 0
\\
\hline
$\left({\bf 6,2},\frac{5}{6} \right)+{\mathrm h.c.}$
&$ 2 m_1 - \frac{\lambda_1 \phi_2}{3 \sqrt{2}} 
- \frac{\lambda_1 \phi_3}{6} 
$
& 10
& 6
& 10
\\
\hline
$\left({\bf \overline{6},2},\frac{1}{6} \right)+{\mathrm h.c.}$
& $ 2 m_1 - \frac{\lambda_1 \phi_2}{3 \sqrt{2}} 
+ \frac{\lambda_1 \phi_3}{6} $
& 10
& 6
& $\frac{2}{5}$
\\
\hline
$\left({\bf 3,3},\frac{2}{3} \right)+{\mathrm h.c.}$
& $ 2 m_1 - \frac{\lambda_1 \phi_1}{\sqrt{6}} 
+ \frac{\lambda_1 \phi_2}{3 \sqrt{2}} $
& 3
& 12
& $\frac{24}{5}$
\\
\hline
$\left({\bf 3,1},\frac{5}{3} \right)+{\mathrm h.c.}$
& $ 2 m_1 + \frac{\lambda_1 \phi_1}{\sqrt{6}} 
+ \frac{\lambda_1 \phi_2}{3 \sqrt{2}} 
- \frac{2\lambda_1 \phi_3}{3}
$
& 1
& 0
& 10
\\
\hline
$\left({\bf 1,3},0 \right)$
& $ 2 m_1 - \frac{\lambda_1 \phi_1}{\sqrt{6}} 
+ \frac{\sqrt{2} \lambda_1 \phi_2}{3} $
& 0
& 2
& 0
\\
\hline
$\left({\bf 1,2},\frac{3}{2} \right)+{\mathrm h.c.}$
& $ 2 m_1 + \frac{\lambda_1 \phi_2}{\sqrt{2}} 
- \frac{\lambda_1 \phi_3}{2} 
$
& 0
& 1
& $\frac{27}{5}$
\\
\end{tabular} 
\end{center}
\end{table}

Putting these values into Eq. (\ref{RGE}), 
the unification condition produces two individual equations,  
\be
\alpha_3 \left(M_G \right)
= \alpha_2 \left(M_G \right),
\label{3=2}
\ee
and
\be
\alpha_3 \left(M_G \right)
= \alpha_1 \left(M_G \right).
\label{3=1}
\ee  
Setting all VEVs at the GUT scale,  
$\phi_1 \sim \phi_2 \sim \phi_3 \sim |v_R| \sim M_G$,  
and the remaining dimensionless coefficients 
of order one, we can search whether Eqs. (\ref{3=2}) and (\ref{3=1}) 
have a solution for $M_G$ below the Planck scale, 
$M_G \leq M_{\mathrm{Planck}}$.
If such a solution exists, 
it would limit the parameters in the superpotential Eq. (\ref{lee}) 
to some restricted region.  
\section{Conclusion}
We find the general formulation for the proton decay rate
in the minimal renormalizable SUSY SO(10) models.
Using this generic formulation one can find whether the minimal SUSY SO(10)
grand unified theory has been excluded.  

Recently, using their Yukawa couplings  
(Eqs. (8) and (9) in Ref. \cite{rabindra2}),
Goh-Mohapatra-Nasri-Ng obtained the allowed region of
$\left(x, y, z\right)$ which correspond to
$\left(\frac{a}{d}, -\frac{b}{d}, -\frac{c}{d}\right)$
in our notation.  However, they did not discuss the concrete form 
of the superpotential and, therefore, compatibilities of 
their superpotential with the other constraints are not clear 
in their paper.  Also, as we have mentioned above, 
there appears a non zero $x$ value even without the ${\bf 54}$ 
dimensional Higgs field.  Further, besides the color triplet Higgs fields, 
there is a much richer Higgs particle contents.  These additional Higgs 
fields may cause a pathology of the gauge couplings unification.  
This paper presents a relationship among these comprehensive 
but tightly connected problems.   
\section*{Acknowledgement}
This work was supported by the Ministry of the Science and Technology 
of the Republic of Croatia under contracts No.0098003 and No.0119261.  
\section*{Note added}
In the recent paper [arXiv:hep-ph/0402122] and the revised version of the paper 
[arXiv:hep-ph/0204097], the authors claimed that they disagree with our results 
for mass matrices.  We point out that our results satisfy all possible 
consistency checks.  
Namely, for arbitrary couplings $m_{1,2,3}$ and $\lambda_{1,2,3,4}$ 
in Eq. (\ref{lee}) there are solutions characterlized by $|v_R| = 0$,  
particularly $SU(5) \times U(1)$, $SU(5) \times U(1) \,\, {\mathrm flipped}$, 
$SU(3)_C \times SU(2)_L \times SU(2)_R \times U(1)_{B-L}$ and 
$SU(3)_C \times SU(2)_L \times U(1)_R \times U(1)_{B-L}$ symmetry preserving vacua.  
(see, Eqs. (\ref{VEV1})--(\ref{VEV3}) with $|v_R| = 0$).  
In all the above four symmetry breakings and in all the mass matrices, 
the field $\Phi$ decouples from the other set of fields, $H$, 
$\overline{\Delta}$ and $\Delta$.  
Moreover, all our mass eigenvalues 
for the multiplets coming from the field $\Phi$, 
coincide with the corresponding results in \cite{he}.  
Furthermore, for the $SU(5)$ symmetry breaking case, 
$|v_R| \neq 0$ (see, the end of Sec.III), 
all our mass eigenvalues (472 in total) and the corresponding 
multiplets under $SU(3)_C \times SU(2)_L \times U(1)_Y$ 
are grouped according to the $SU(5)$ irreducible representations 
with the correct would-be NG fields.  
The details will be published in a separate publication.  
\section*{Appendix}
\subsection*{A. Dimension-five operators}
In this appendix, we list the explicit form of the various 
interaction coefficients.  

We use the following notations for the mixing matrices 
which diagonalize the squark, slepton mass-squared matrices and 
chargino, neutralino mass matrices.  
Squark, slepton mass-squared matrix $M_{\widetilde{f}}^2$,  
chargino and neutralino mass matrices $M_C$ and $M_N$ are diagonalized 
by the unitary matrices $U_{\widetilde f}$, $O_L$, $O_R$ and $O_N$, 
respectively.  
$$
U_{\widetilde f} \, M_{\widetilde{f}}^2 \, U_{\widetilde f}^\dagger 
=
{\mathrm{diag}} 
(m_{\widetilde{f}_{1}}^2,....,m_{\widetilde{f}_{6}}^2),
\nonumber
$$ $$
O_R \, M_C \, O_L^{\dagger} 
=
{\mathrm{diag}} 
(m_{\widetilde{\chi}_{1}^{-}}, m_{\widetilde{\chi}_{2}^{-}}),
\nonumber
$$ $$
O_N^* \, M_N \, O_N^{\dagger} 
=
{\mathrm{diag}} 
(m_{\widetilde{\chi}_{1}^{0}}, m_{\widetilde{\chi}_{2}^{0}}, 
m_{\widetilde{\chi}_{3}^{0}}, m_{\widetilde{\chi}_{4}^{0}}).
\nonumber
\eqno{(A.1)}
$$
For the dimension-five operators, 
we have the following expressions 
\footnotemark
\footnotetext{We use a notation for an anti-symmetric tensor, 
$A^{[ijk]l} \equiv A^{ijkl}-A^{kjil}.$}  \\
$$
C_{L}^{(\tilde{u} \tilde{d} u e) XYij}
\equiv
C_L^{[ijk]l}
(U_{\widetilde{u}}^{*})_{Xk} (U_{\widetilde{d}}^{*})_{Yl},
\eqno{(A.2)} 
$$ $$
C_{L}^{(\tilde{u} \tilde{u} d e) XYij}
\equiv
C_L^{[kjl]m}
(U_{\widetilde{u}}^{*})_{Xk} (U_{\widetilde{u}}^{*})_{Yl}
(V_{\mathrm{CKM}})_{im},
\eqno{(A.3)}
$$ $$
C_{R}^{(\tilde{u} \tilde{d} u e) XYij}
\equiv
(C_R^{*klji}-C_R^{*iljk})
(U_{\widetilde{u}}^{*})_{X,k+3} (U_{\widetilde{d}}^{*})_{Y,l+3},
\eqno{(A.4)}
$$ $$
C_{R}^{(\tilde{u} \tilde{u} d e) XYij}
\equiv
(C_R^{*klji}-C_R^{*iljk})
(U_{\widetilde{u}}^{*})_{X,k+3} (U_{\widetilde{u}}^{*})_{Y,l+3},
\eqno{(A.5)}
$$ $$ 
C_{L}^{(\tilde{u} \tilde{d} d \nu) XYij}
\equiv
(C_L^{mnkl}-C_L^{lknm})
(U_{\widetilde{u}}^{*})_{Xk} (U_{\widetilde{d}}^{*})_{Yl}
(V_{\mathrm{CKM}})_{im} (U_{\mathrm{MNS}})_{jn},
\eqno{(A.6)}
$$ $$ 
C_{L}^{(\tilde{d} \tilde{d} u \nu) XYij}
\equiv
(C_L^{lnik}-C_L^{knil})
(U_{\widetilde{d}}^{*})_{Xk} (U_{\widetilde{d}}^{*})_{Yl}
(U_{\mathrm{MNS}})_{jn},
\eqno{(A.7)}
$$ $$
C_{L}^{(\tilde{u} \tilde{e} u d) XYij}
\equiv
C_L^{[kli]m}
(U_{\widetilde{u}}^{*})_{Xk} (U_{\widetilde{e}}^{*})_{Yl}
(V_{\mathrm{CKM}})_{jm},
\eqno{(A.8)}
$$ $$
C_{L}^{(\tilde{d} \tilde{e} u u) XYij}
\equiv
C_L^{[ilj]k}
(U_{\widetilde{d}}^{*})_{Xk} (U_{\widetilde{e}}^{*})_{Yl},
\eqno{(A.9)}
$$ $$
C_{R}^{(\tilde{u} \tilde{e} u d) XYij}
\equiv
(C_R^{*jkli}-C_R^{*kjli})
(U_{\widetilde{u}}^{*})_{X,k+3} (U_{\widetilde{e}}^{*})_{Y,l+3},
\eqno{(A.10)}
$$ $$
C_{R}^{(\tilde{d} \tilde{e} u u) XYij}
\equiv
(C_R^{*jkli}-C_R^{*iklj})
(U_{\widetilde{d}}^{*})_{X,k+3} (U_{\widetilde{e}}^{*})_{Y,l+3},
\eqno{(A.11)}
$$ $$
C_{L}^{(\tilde{d} \tilde{\nu} u d) XYij}
\equiv
(C_L^{klim}-C_L^{mlik})
(U_{\widetilde{d}}^{*})_{Xk} (U_{\widetilde{\nu}}^{*})_{Yl}
(V_{\mathrm{CKM}})_{jm},
\eqno{(A.12)}
$$ $$
C_{L}^{(\tilde{u} \tilde{\nu} d d) XYij}
\equiv
(C_L^{nlkm}-C_L^{mlkn})
(U_{\widetilde{u}}^{*})_{Xk} (U_{\widetilde{\nu}}^{*})_{Yl}
(V_{\mathrm{CKM}})_{im} (V_{\mathrm{CKM}})_{jn}.
\eqno{(A.13)}
$$
In Eqs. (A.6) and (A.7), 
it should be noticed that the neutrinos in the final states should 
be rotated from the flavor eigenstates to the mass eigenstates 
by using the Maki-Nakagawa-Sakata (MNS) mixing 
matrix \cite{MNS}, $U_{\mathrm{MNS}}$.  
\subsection*{B. Sparticles interactions}
We use the following notations for the 
quark-gluino-squark, quark(lepton)-chargino-squark(slepton) 
and quark(lepton)-neutralino-squark(slepton) interactions,
\begin{itemize}
\item quark-gluino-squark interactions \\
$$
{\mathcal L}_{int}
=
-i \sqrt{2} u^c_i
\left[G_{iX}^{L(u)}P_{L}+G_{iX}^{R(u)}P_{R} \right]
\widetilde{g} \widetilde{u}_{X} 
-i 
\sqrt{2} d^c_i
\left[G_{iX}^{L(d)}P_{L}+G_{iX}^{R(d)}P_{R} \right]
\widetilde{g} \widetilde{d}_{X} +\mathrm{h.c.}
\eqno{(B.1)}
$$
\item quark(lepton)-chargino-squark(slepton) interactions \\
\bea
{\mathcal L}_{int}
&=&
u^c_i
\left[C_{iAX}^{L(u)}P_{L}+C_{iAX}^{R(u)}P_{R} \right]
\widetilde{\chi}_A^{+} \widetilde{d}_{X}
+
d^c_i
\left[C_{iAX}^{L(d)}P_{L}+C_{iAX}^{R(d)}P_{R} \right]
\widetilde{\chi}_A^{+} \widetilde{u}_{X}
\nonumber
\eea
$$
+
\nu^c_i C_{iAX}^{R(\nu)}P_{R}
\widetilde{\chi}_A^{+} \widetilde{e}_{X}
+
e^c_i
\left[C_{iAX}^{L(e)}P_{L}+C_{iAX}^{R(e)}P_{R} \right]
\widetilde{\chi}_A^{+} \widetilde{\nu}_{X} +\mathrm{h.c.}
\eqno{(B.2)}
$$
\item quark(lepton)-neutralino-squark(slepton) interactions \\
\bea
{\mathcal L}_{int}
&=&
u^c_i
\left[N_{iAX}^{L(u)}P_{L}+N_{iAX}^{R(u)}P_{R} \right]
\widetilde{\chi}_A^{0} \widetilde{u}_{X}
+
d^c_i
\left[N_{iAX}^{L(d)}P_{L}+N_{iAX}^{R(d)}P_{R} \right]
\widetilde{\chi}_A^{0} \widetilde{d}_{X}
\nonumber
\eea
$$
+
\nu^c_i N_{iAX}^{R(\nu)}P_{R}
\widetilde{\chi}_A^{0} \widetilde{\nu}_{X}
+
e^c_i
\left[N_{iAX}^{L(e)}P_{L}+N_{iAX}^{R(e)}P_{R} \right]
\widetilde{\chi}_A^{0} \widetilde{e}_{X}  +\mathrm{h.c.}
\eqno{(B.3)}
$$
\end{itemize}
Explicitly, we have the following expressions
$$
G_{iX}^{L(u)} \equiv
g_3 (U_{\tilde{u}}^{*})_{X,i+3},
\eqno{(B.4)}
$$ $$
G_{iX}^{R(u)} \equiv 
g_3 (U_{\tilde{u}}^{*})_{Xi},
\eqno{(B.5)}
$$ $$
G_{iX}^{L(d)} \equiv
g_3 (U_{\tilde{d}}^{*})_{X,i+3},
\eqno{(B.6)}
$$ $$
G_{iX}^{R(d)} \equiv
g_3 (U_{\tilde{d}}^{*})_{Xk} (V_{\mathrm{CKM}}^{*})_{ik}, 
\eqno{(B.7)}
$$ $$
C_{iAX}^{L(u)} \equiv
g \frac{m_{u_i}}{\sqrt{2} M_{\mathrm{W}} \sin \beta} 
(O_{R}^*)_{A 2} (U_{\tilde d}^*)_{Xi} , 
\eqno{(B.8)}
$$ $$
C_{iAX}^{R(u)} \equiv 
g \left\{-(O_{L}^*)_{A1} (U_{\tilde d}^*)_{Xi}
+
\frac{m_{d_i}}{\sqrt{2} M_{\mathrm{W}} \cos \beta} 
(O_{L}^*)_{A 2} (U_{\tilde u}^*)_{X,i+3} \right\} , 
\eqno{(B.9)}
$$ $$
C_{iAX}^{L(d)} \equiv
g \frac{m_{d_i}}{\sqrt{2} M_{\mathrm{W}} \cos \beta} 
(O_{L}^*)_{A 2} (U_{\tilde u}^*)_{Xi} , 
\eqno{(B.10)}
$$ $$
C_{iAX}^{R(d)} \equiv 
g \left\{-(O_{R}^*)_{A1} (U_{\tilde u}^*)_{Xk}
+
\frac{m_{u_k}}{\sqrt{2} M_{\mathrm{W}} \sin \beta} 
(O_{R}^*)_{A 2} (U_{\tilde u}^*)_{X,k+3} \right\} 
(V_{\mathrm{CKM}}^{*})_{ik},
\eqno{(B.11)}
$$ $$
C_{iAX}^{R(\nu)} \equiv 
g \left\{-(O_{L}^*)_{A1} (U_{\tilde d}^*)_{Xk}
+\frac{m_{e_k}}{\sqrt{2} M_{\mathrm{W}} \cos \beta} 
(O_{L}^*)_{A 2} (U_{\tilde \nu}^*)_{X,k+3} \right\} 
(U_{\mathrm{MNS}}^{*})_{ik}, 
\eqno{(B.12)}
$$ $$
C_{iAX}^{L(e)} \equiv
g \frac{m_{e_i}}{\sqrt{2} M_{\mathrm{W}} \cos \beta} 
(O_{L}^*)_{A 2} (U_{\tilde \nu}^*)_{Xi} , 
\eqno{(B.13)}
$$ $$
C_{iAX}^{R(e)} \equiv
- g \left\{-(O_{R}^*)_{A1} (U_{\tilde \nu}^*)_{Xk}
+
\frac{m_{u_k}}{\sqrt{2} M_{\mathrm{W}} \sin \beta} 
(O_{R}^*)_{A2} (U_{\tilde u}^*)_{X,k+3} \right\} 
(V_{\mathrm{CKM}}^{*})_{ik}, 
\eqno{(B.14)}
$$ $$
N_{iAX}^{L(u)} \equiv 
-\frac{g}{\sqrt{2}} \left\{
\frac{m_{u_i}}{M_{\mathrm{W}} \sin \beta} 
(O_{N}^*)_{A4} (U_{\tilde u}^*)_{Xi} 
-
\frac{4}{3} \tan \theta_{W} (O_{N}^{*})_{A1} 
(U_{\tilde u}^{*})_{X,i+3} \right\}, 
\eqno{(B.15)}
$$ $$
N_{iAX}^{R(u)} \equiv
-\frac{g}{\sqrt{2}} \left\{
\frac{m_{u_i}}{M_{\mathrm{W}} \sin \beta} 
(O_{N}^*)_{A4} (U_{\tilde u}^*)_{X,i+3} 
+
\left[(O_{N}^{*})_{A2} + 
\frac{1}{3}\tan \theta_{W} (O_{N}^{*})_{A1} \right] 
(U_{\tilde u}^{*})_{Xi} \right\}, 
\eqno{(B.16)}
$$ $$
N_{iAX}^{L(d)} \equiv
-\frac{g}{\sqrt{2}} \left\{
\frac{m_{d_i}}{M_{\mathrm{W}} \cos \beta} 
(O_{N}^*)_{A3} (U_{\tilde d}^*)_{Xi} 
+
\frac{2}{3} \tan \theta_{W} (O_{N}^{*})_{A1} 
(U_{\tilde d}^{*})_{X,i+3} \right\}, 
\eqno{(B.17)}
$$ $$
N_{iAX}^{R(d)} \equiv
-\frac{g}{\sqrt{2}} \left\{
\frac{m_{d_k}}{M_{\mathrm{W}} \cos \beta} 
(O_{N}^*)_{A3} (U_{\tilde d}^*)_{X,k+3} 
+
\left[-(O_{N}^{*})_{A2}+
\frac{1}{3} \tan \theta_{W} (O_{N}^{*})_{A1} \right]
(U_{\tilde d}^{*})_{Xk} \right\}
(V_{\mathrm{CKM}}^{*})_{ik}, 
\eqno{(B.18)}
$$ $$
N_{iAX}^{R(\nu)} \equiv
-\frac{g}{\sqrt{2}} 
\left[(O_{N}^{*})_{A2} - 
\tan \theta_{W} (O_{N}^{*})_{A1} \right] 
(U_{\tilde \nu}^{*})_{X,k}
(U_{\mathrm{MNS}}^{*})_{ik}, 
\eqno{(B.19)}
$$ $$
N_{iAX}^{L(e)} \equiv
-\frac{g}{\sqrt{2}} \left\{
\frac{m_{e_i}}{M_{\mathrm{W}} \cos \beta} 
(O_{N}^*)_{A3} (U_{\tilde e}^*)_{Xi}
+
\frac{2}{3} \tan \theta_{W} (O_{N}^{*})_{A1} 
(U_{\tilde e}^{*})_{X,i+3} \right\}, 
\eqno{(B.20)}
$$ $$
N_{iAX}^{R(e)} \equiv
-\frac{g}{\sqrt{2}} \left\{
\frac{m_{e_i}}{M_{\mathrm{W}} \cos \beta} 
(O_{N}^*)_{A3} (U_{\tilde e}^*)_{X,i+3} 
+
\left[-(O_{N}^{*})_{A2}+
\frac{1}{3} \tan \theta_{W} (O_{N}^{*})_{A1} \right]
(U_{\tilde e}^{*})_{Xi} \right\}.
\eqno{(B.21)}
$$
These expressions are found in \cite{gunion}, but only 
for the quark sector.  So here we write them explicitly.  
\subsection*{C. Dimension-six operators}
For the dimension-six operator, we devide the coefficients 
into three parts according to the dressed sparticles,  
$$
C_{LL}^{(udue)ij}
=C_{LL}^{(udue)ij}(\widetilde{g})
+C_{LL}^{(udue)ij}(\widetilde{\chi}^{0})
+C_{LL}^{(udue)ij}(\widetilde{\chi}^{\pm}),
\eqno{(C.1)}
$$
etc.  
Then we have the following expressions. These expressions 
have the same forms as \cite{Goto}.  However, ours are 
different from them in the neutrino sector as was mentioned 
in the end of Appendix. A.   
$$
C_{LL}^{(udue)ij}(\widetilde{g})
=\frac{4}{3}\frac{1}{m_{\widetilde{g}}}
C_L^{(udue)XY1j}G_{1X}^{R(u)}G_{iY}^{R(d)}
F\left(\frac{m_{\widetilde{g}}^2}{m_{\widetilde{u}_{X}}^2}, 
\frac{m_{\widetilde{g}}^2}{m_{\widetilde{d}_{Y}}^2} \right),
\eqno{(C.2)}
$$ 
\bea 
C_{LL}^{(udue)ij}(\widetilde{\chi}^{\pm})
&=&\frac{1}{m_{\widetilde{\chi}_A^{+}}}
\left[-C_L^{(udue)XY1j}C_{1AY}^{R(u)}C_{iAX}^{R(d)}
F\left(\frac{m_{\widetilde{\chi}_A^{+}}^2}{m_{\widetilde{u}_{X}}^2}, 
\frac{m_{\widetilde{\chi}_A^{+}}^2}{m_{\widetilde{d}_{Y}}^2} \right)
\right.
\nonumber
\eea
$$
+\left. C_L^{(d\nu ud)XY1i}C_{1AX}^{R(d)}C_{jAY}^{R(e)}
F\left(\frac{m_{\widetilde{\chi}_A^{+}}^2}{m_{\widetilde{d}_{X}}^2}, 
\frac{m_{\widetilde{\chi}_A^{+}}^2}{m_{\widetilde{\nu}_{Y}}^2} \right)
\right],
\eqno{(C.3)}
$$ 
\bea
C_{LL}^{(udue)ij}(\widetilde{\chi}^{0})
&=&\frac{1}{m_{\widetilde{\chi}_A^{0}}}
\left[C_L^{(udue)XY1j}N_{1AX}^{R(u)}N_{iAY}^{R(d)}
F\left(\frac{m_{\widetilde{\chi}_A^{0}}^2}{m_{\widetilde{u}_{X}}^2}, 
\frac{m_{\widetilde{\chi}_A^{0}}^2}{m_{\widetilde{d}_{Y}}^2} \right)
\right.
\nonumber
\eea
$$
+
\left.
C_L^{(ueud)XY1i}N_{1AX}^{R(d)}N_{jAY}^{R(e)}
F\left(\frac{m_{\widetilde{\chi}_A^{0}}^2}{m_{\widetilde{d}_{X}}^2}, 
\frac{m_{\widetilde{\chi}_A^{0}}^2}{m_{\widetilde{\nu}_{Y}}^2}\right)
\right],
\eqno{(C.4)}
$$ $$
C_{RL}^{(udue)ij}(\widetilde{g})
=\frac{4}{3}\frac{1}{m_{\widetilde{g}}}
C_L^{(udue)XY1j}G_{1X}^{L(u)}G_{iY}^{L(d)}
F\left(\frac{m_{\widetilde{g}}^2}{m_{\widetilde{u_{X}}}^2}, 
\frac{m_{\widetilde{g}}^2}{m_{\widetilde{d_{Y}}}^2} \right),
\eqno{(C.5)}
$$ $$
C_{RL}^{(udue)ij}(\widetilde{\chi}^{\pm})
=-\frac{1}{m_{\widetilde{\chi}_A^{+}}}
C_L^{(udue)XY1j}C_{1AY}^{L(u)}C_{iAX}^{L(d)}
F\left(\frac{m_{\widetilde{\chi}_A^{+}}^2}{m_{\widetilde{u}_{X}}^2}, 
\frac{m_{\widetilde{\chi}_A^{+}}^2}{m_{\widetilde{d}_{Y}}^2} \right),
\eqno{(C.6)}
$$ \bea
C_{RL}^{(udue)ij}(\widetilde{\chi}^{0})
&=&\frac{1}{m_{\widetilde{\chi}_A^{0}}}
\left[C_L^{(udue)XY1j}N_{1AX}^{L(u)}N_{iAY}^{L(d)}
F\left(\frac{m_{\widetilde{\chi}_A^{0}}^2}{m_{\widetilde{u}_{X}}^2}, 
\frac{m_{\widetilde{\chi}_A^{0}}^2}{m_{\widetilde{d}_{Y}}^2} \right)
\right.
\nonumber
\eea
$$
+
\left.
C_R^{(ueud)XY1i}N_{1AX}^{R(d)}N_{jAY}^{R(e)}
F\left(\frac{m_{\widetilde{\chi}_A^{0}}^2}{m_{\widetilde{u}_{X}}^2}, 
\frac{m_{\widetilde{\chi}_A^{0}}^2}{m_{\widetilde{e}_{Y}}^2}\right)
\right],
\eqno{(C.7)}
$$ $$
C_{LR}^{(udue)ij}(\widetilde{g})
=\frac{4}{3}\frac{1}{m_{\widetilde{g}}}
C_R^{(udue)XY1j}G_{1X}^{R(u)}G_{iY}^{R(d)}
F\left(\frac{m_{\widetilde{g}}^2}{m_{\widetilde{u}_{X}}^2}, 
\frac{m_{\widetilde{g}}^2}{m_{\widetilde{d}_{Y}}^2} \right),
\eqno{(C.8)}
$$ \bea
C_{LR}^{(udue)ij}(\widetilde{\chi}^{\pm})
&=&\frac{1}{m_{\widetilde{\chi}_A^{+}}}
\left[-C_R^{(udue)XY1j}C_{1AY}^{R(u)}C_{iAX}^{R(d)}
F\left(\frac{m_{\widetilde{\chi}_A^{+}}^2}{m_{\widetilde{u}_{X}}^2}, 
\frac{m_{\widetilde{\chi}_A^{+}}^2}{m_{\widetilde{d}_{Y}}^2} \right)
\right.
\nonumber
\eea
$$
+
\left.
C_L^{(d\nu ud)XY1i}C_{1AX}^{L(d)}C_{jAY}^{L(e)}
F\left(\frac{m_{\widetilde{\chi}_A^{+}}^2}{m_{\widetilde{d}_{X}}^2}, 
\frac{m_{\widetilde{\chi}_A^{+}}^2}{m_{\widetilde{\nu}_{Y}}^2}\right)
\right],
\eqno{(C.9)}
$$
\bea
C_{LR}^{(udue)ij}(\widetilde{\chi}^{0})
&=&\frac{1}{m_{\widetilde{\chi}_A^{0}}}
\left[C_R^{(udue)XY1j}N_{1AX}^{R(u)}N_{iAY}^{R(d)}
F\left(\frac{m_{\widetilde{\chi}_A^{0}}^2}{m_{\widetilde{u}_{X}}^2}, 
\frac{m_{\widetilde{\chi}_A^{0}}^2}{m_{\widetilde{d}_{Y}}^2} \right)
\right.
\nonumber \eea
$$
+
\left.
C_L^{(ueud)XY1i}N_{1AX}^{L(d)}N_{jAY}^{L(e)}
F\left(\frac{m_{\widetilde{\chi}_A^{0}}^2}{m_{\widetilde{d}_{X}}^2}, 
\frac{m_{\widetilde{\chi}_A^{0}}^2}{m_{\widetilde{e}_{Y}}^2}\right)
\right],
\eqno{(C.10)}
$$ $$
C_{RR}^{(udue)ij}(\widetilde{g})
=\frac{4}{3}\frac{1}{m_{\widetilde{g}}}
C_R^{(udue)XY1j}G_{1X}^{L(u)}G_{iY}^{L(d)}
F\left(\frac{m_{\widetilde{g}}^2}{m_{\widetilde{u_{X}}}^2}, 
\frac{m_{\widetilde{g}}^2}{m_{\widetilde{d}_{Y}}^2} \right),
\eqno{(C.11)}
$$ $$
C_{RR}^{(udue)ij}(\widetilde{\chi}^{\pm})
=-\frac{1}{m_{\widetilde{\chi}_A^{+}}}
C_R^{(udue)XY1j}C_{1AY}^{L(u)}C_{iAX}^{L(d)}
F\left(\frac{m_{\widetilde{\chi}_A^{+}}^2}{m_{\widetilde{u}_{X}}^2}, 
\frac{m_{\widetilde{\chi}_A^{+}}^2}{m_{\widetilde{d}_{Y}}^2} \right),
\eqno{(C.12)}
$$ \bea
C_{RR}^{(udue)ij}(\widetilde{\chi}^{0})
&=&\frac{1}{m_{\widetilde{\chi}_A^{0}}}
\left[C_L^{(udue)XY1j}N_{1AX}^{L(u)}N_{iAY}^{L(d)}
F\left(\frac{m_{\widetilde{\chi}_A^{0}}^2}{m_{\widetilde{u}_{X}}^2}, 
\frac{m_{\widetilde{\chi}_A^{0}}^2}{m_{\widetilde{d}_{Y}}^2} \right)
\right.
\nonumber \eea
$$
+
\left.
C_R^{(ueud)XY1i}N_{1AX}^{R(d)}N_{jAY}^{R(e)}
F\left(\frac{m_{\widetilde{\chi}_A^{0}}^2}{m_{\widetilde{u}_{X}}^2}, 
\frac{m_{\widetilde{\chi}_A^{0}}^2}{m_{\widetilde{e}_{Y}}^2}\right)
\right],
\eqno{(C.13)}
$$ \bea
C_{LL}^{(udd\nu)ijk}(\widetilde{g})
&=&\frac{4}{3}\frac{1}{m_{\widetilde{g}}}
\left[C_L^{(udd\nu)XYjk}G_{1X}^{R(u)}G_{iY}^{R(d)}
F\left(\frac{m_{\widetilde{g}}^2}{m_{\widetilde{u}_{X}}^2}, 
\frac{m_{\widetilde{g}}^2}{m_{\widetilde{d}_{Y}}^2} \right)
\right.
\nonumber \eea
$$
+
\left.
C_L^{(ddu\nu)XY1k}G_{jX}^{R(d)}G_{iY}^{R(d)}
F\left(\frac{m_{\widetilde{g}}^2}{m_{\widetilde{d}_{X}}^2}, 
\frac{m_{\widetilde{g}}^2}{m_{\widetilde{d}_{Y}}^2} \right)
\right],
\eqno{(C.14)}
$$ \bea
C_{LL}^{(udd\nu)ijk}(\widetilde{\chi}^{\pm})
&=&\frac{1}{m_{\widetilde{\chi}_A^{+}}}
\left[-C_L^{(udd\nu)XYjk}C_{1AY}^{R(u)}C_{iAX}^{R(d)}
F\left(\frac{m_{\widetilde{\chi}_A^{+}}^2}{m_{\widetilde{u}_{X}}^2}, 
\frac{m_{\widetilde{\chi}_A^{+}}^2}{m_{\widetilde{d}_{Y}}^2} \right)
\right.
\nonumber \eea
$$
+
\left.
C_L^{(ueud)XY1i}C_{jAX}^{R(u)}C_{kAY}^{R(e)}
F\left(\frac{m_{\widetilde{\chi}_A^{+}}^2}{m_{\widetilde{d}_{X}}^2}, 
\frac{m_{\widetilde{\chi}_A^{+}}^2}{m_{\widetilde{\nu}_{Y}}^2}\right)
\right],
\eqno{(C.15)}
$$ \bea
C_{LL}^{(udd\nu)ijk}(\widetilde{\chi}^{0})
&=&\frac{1}{m_{\widetilde{\chi}_A^{0}}}
\left[C_L^{(udd\nu)XYjk}N_{1AX}^{R(u)}N_{iAY}^{R(d)}
F\left(\frac{m_{\widetilde{\chi}_A^{0}}^2}{m_{\widetilde{u}_{X}}^2}, 
\frac{m_{\widetilde{\chi}_A^{0}}^2}{m_{\widetilde{d}_{Y}}^2} \right)
\right.
\nonumber\\
&+&
C_L^{(ddu\nu)XY1k}N_{jAX}^{R(d)}N_{iAY}^{R(e)}
F\left(\frac{m_{\widetilde{\chi}_A^{0}}^2}{m_{\widetilde{d}_{X}}^2}, 
\frac{m_{\widetilde{\chi}_A^{0}}^2}{m_{\widetilde{d}_{Y}}^2}\right)
\nonumber\\
&+&
C_L^{(d\nu ud)XY1i}N_{jAX}^{R(d)}N_{kAY}^{R(e)}
F\left(\frac{m_{\widetilde{\chi}_A^{0}}^2}{m_{\widetilde{d}_{X}}^2}, 
\frac{m_{\widetilde{\chi}_A^{0}}^2}{m_{\widetilde{\nu}_{Y}}^2}\right)
\nonumber \eea
$$
+
\left.
C_L^{(u\nu dd)XYji}N_{1AX}^{R(u)}N_{kAY}^{R(\nu)}
F\left(\frac{m_{\widetilde{\chi}_A^{0}}^2}{m_{\widetilde{u}_{X}}^2}, 
\frac{m_{\widetilde{\chi}_A^{0}}^2}{m_{\widetilde{\nu}_{Y}}^2}\right)
\right],
\eqno{(C.16)}
$$
$$
C_{RL}^{(udd\nu)ijk}(\widetilde{g})
=\frac{4}{3}\frac{1}{m_{\widetilde{g}}}
C_L^{(udd\nu)XYjk}G_{1X}^{L(u)}G_{iY}^{L(d)}
F\left(\frac{m_{\widetilde{g}}^2}{m_{\widetilde{u}_{X}}^2}, 
\frac{m_{\widetilde{g}}^2}{m_{\widetilde{d}_{Y}}^2} \right),
\eqno{(C.17)}
$$ \bea
C_{RL}^{(udd\nu)ijk}(\widetilde{\chi}^{\pm})
&=&\frac{1}{m_{\widetilde{\chi}_A^{+}}}
\left[-C_L^{(udd\nu)XYjk}C_{1AY}^{L(u)}C_{iAX}^{L(d)}
F\left(\frac{m_{\widetilde{\chi}_A^{+}}^2}{m_{\widetilde{u}_{X}}^2}, 
\frac{m_{\widetilde{\chi}_A^{+}}^2}{m_{\widetilde{d}_{Y}}^2} \right)
\right.
\nonumber \eea
$$
+
\left.
C_R^{(ueud)XY1i}C_{jAX}^{R(u)}C_{kAY}^{R(e)}
F\left(\frac{m_{\widetilde{\chi}_A^{+}}^2}{m_{\widetilde{d}_{X}}^2}, 
\frac{m_{\widetilde{\chi}_A^{+}}^2}{m_{\widetilde{\nu}_{Y}}^2}\right)
\right],
\eqno{(C.18)}
$$ $$
C_{RL}^{(udd\nu)ijk}(\widetilde{\chi}^{0})
=\frac{1}{m_{\widetilde{\chi}_A^{0}}}
C_L^{(udd\nu)XYjk}N_{1AX}^{L(u)}N_{iAY}^{L(d)}
F\left(\frac{m_{\widetilde{\chi}_A^{0}}^2}{m_{\widetilde{u}_{X}}^2}, 
\frac{m_{\widetilde{\chi}_A^{0}}^2}{m_{\widetilde{d}_{Y}}^2} \right),
\eqno{(C.19)}
$$ $$
C_{RL}^{(ddu\nu)ijk}(\widetilde{g})
=\frac{4}{3}\frac{1}{m_{\widetilde{g}}}
C_L^{(udd\nu)XY1k}G_{iX}^{L(d)}G_{jY}^{L(d)}
F\left(\frac{m_{\widetilde{g}}^2}{m_{\widetilde{d}_{X}}^2}, 
\frac{m_{\widetilde{g}}^2}{m_{\widetilde{d}_{Y}}^2} \right),
\eqno{(C.20)}
$$ $$
C_{RL}^{(ddu\nu)ijk}(\widetilde{\chi}^{\pm})
=0,
\eqno{(C.21)}
$$ $$
C_{RL}^{(ddu\nu)ijk}(\widetilde{\chi}^{0})
=\frac{1}{m_{\widetilde{\chi}_A^{0}}}
C_L^{(ddu\nu)XY1k}N_{iAX}^{L(d)}N_{jAY}^{L(d)}
F\left(\frac{m_{\widetilde{\chi}_A^{0}}^2}{m_{\widetilde{d}_{X}}^2}, 
\frac{m_{\widetilde{\chi}_A^{0}}^2}{m_{\widetilde{d}_{Y}}^2} \right).
\eqno{(C.22)}
$$
Here we have defined a loop function,
$$
F(x,y) \equiv \frac{x \, y}{x-y} 
\left(\frac{1}{1-x}\log x-\frac{1}{1-y}\log y \right).
\eqno{(C.23)}
$$

\end{document}